\documentclass[aps,pre,showpacs,twocolumn]{revtex4}
\usepackage{graphicx}
\usepackage{graphicx}
\usepackage{amsmath}
\usepackage{amsfonts}
\usepackage{xcolor}
\begin{document}
\title{Breathing and switching cyclops states in Kuramoto networks with higher-mode coupling}

%\title{Higher-order coupling induces Cyclope states in repulsive Kuramoto networks}

%\title{Cyclope states in repulsive Kuramoto networks: the role of high-order coupling}
%\title{Anti-resonance in switching systems}

\author{Maxim I. Bolotov$^{1}$, Vyacheslav O. Munyayev$^{1}$, Lev A. Smirnov$^{1}$, Grigory V. Osipov$^{1}$, and Igor Belykh$^{2}$\footnote{Corresponding author, e-mail: ibelykh@gsu.edu}}

\address{$^1$Department of Control Theory, Lobachevsky State University of Nizhny Novgorod,
	23 Gagarin Avenue, Nizhny Novgorod, 603022, Russia\\
	$^2$Department of Mathematics and Statistics and Neuroscience Institute, Georgia State University, P.O. Box 4110, Atlanta, Georgia, 30302-410, USA}	
\begin{abstract}
Cyclops states are intriguing cluster patterns observed in oscillator networks, including neuronal ensembles. 
The concept of cyclops states formed by two distinct, coherent clusters and a solitary oscillator was introduced in [Munyayev {\it et al.}, Phys. Rev. Lett. 130, 107021 (2023)], where we explored the surprising prevalence of such states in repulsive Kuramoto networks of rotators with higher-mode harmonics in the coupling. This paper extends our analysis to understand the mechanisms responsible for destroying the cyclops' states and inducing new dynamical patterns called breathing and switching cyclops' states. We first analytically study the existence and stability of cyclops states in the Kuramoto-Sakaguchi networks of two-dimensional oscillators with inertia as a function of the second coupling harmonic. We then describe two bifurcation scenarios that give birth to breathing and switching cyclops states. We demonstrate that these states and their hybrids are prevalent across a wide coupling range and are robust against a relatively large intrinsic frequency detuning. Beyond the Kuramoto networks, breathing and switching cyclops states promise to strongly manifest in other physical and biological networks, including coupled theta-neurons.
\end{abstract}

\date{\today}
\draft \maketitle

\section{Introduction} 
Phase oscillator networks have emerged as a paradigmatic reduced model for describing emergent cooperative properties of diverse real-world systems, including neuronal networks \cite{rinzel1998analysis,ermentrout2001traveling,hoppensteadt2012weakly},  laser arrays  \cite{kozyreff2000global,ding2019dispersive,nair2021using}, and power grids \cite{motter2013spontaneous,dorfler2013synchronization,berner2021adaptive}.
The celebrated Kuramoto model of one-dimensional oscillators (1D) \cite{kuramoto1975self, strogatz2000kuramoto} and its extension to two-dimensional (2D) oscillators with inertia  \cite{ermentrout} captures the essence of the phase networks and provides a concise framework to explore the richness of their cooperative dynamics \cite{acebron,barreto2008synchronization, ott2008low,hong2007entrainment, pikovsky2008partially, maistrenko2004mechanism,dorfler2011critical}. These dynamics include full \cite{tanaka1997first,tanaka1997self,ji2014low,munyaev2020analytical,komarov2014synchronization}, partial  \cite{martens2009exact,barabash2021partial}, explosive \cite{gomez2011explosive,ji2013cluster,skardal2014disorder} and asymmetry-induced  synchronization \cite{nishikawa2016symmetric,nicolaou2019multifaceted}, 
chimeras  \cite{kuramoto2002coexistence, abrams2004chimera, abrams2008solvable, panaggio2015chimera, zakharova2014chimera,panaggio2015chimera,olmi2015intermittent,bolotov2016marginal,bolotov2018simple}, solitary states  \cite{jaros2015chimera,maistrenko2017smallest,jaros2018solitary,teichmann2019solitary,munyayev2022stability}, clusters \cite{olmi2014hysteretic,belykh2016bistability,brister2020three,ronge2021splay}, generalized splay \cite{berner2021generalized} and cyclops states \cite{munyayev2023cyclops}. The cooperative dynamics of Kuramoto networks with attractive coupling have been studied more extensively than their counterparts in repulsive networks. While full synchronization is the simplest and most dominant rhythm in attractive networks, splay \cite{tsimring2005repulsive,gao2019repulsive}, generalized and cluster splay states \cite{berner2021generalized,ronge2021splay} are expected to be the most probable patterns in repulsive networks. Yet, a complete understanding of rhythmogenesis in repulsive networks is still lacking.
Two repulsively coupled oscillators tend to achieve anti-phase synchronization; however, predicting an outcome of such interactions in large repulsive networks is often elusive.  Notably, the role of repulsive connections can be counterintuitive, especially in networks with mixed attractive and repulsive coupling \cite{belykh2008weak,nishikawa2010network,belykh2015synergistic,reimbayev2017two}. For example, the addition of pairwise repulsive inhibition to excitatory networks of bursting neurons can induce synchronization to contrast with one's expectations \cite{belykh2015synergistic}. 

In the context of Kuramoto-type networks, the prevailing approach is to model interactions by the first sinusoidal harmonic from a Fourier decomposition of a $2\pi$-periodic coupling function. This simplest choice of the coupling form adequately describes many dynamical features of real-world networks and is analytically tractable. However, higher-order coupling harmonics have been observed to play a significant role in 
rhythmogenesis in various scenarios. These encompass  Kuramoto-type models of neuronal plasticity
\cite{seliger2002plasticity,niyogi2009learning}, coupled electrochemical oscillators \cite{kiss2005predicting}, and Josephson junctions \cite{goldobin2013phase}. In particular, previous research has demonstrated that augmenting the classical Kuramoto model with higher-mode coupling can result in the emergence of multiple phase-locked states  \cite{komarov2013multiplicity,berner2023synchronization} and facilitate switching between synchrony clusters \cite{skardal2011cluster}.

In a recent work \cite{munyayev2023cyclops}, we 
studied rhythmogenesis in repulsive Kuramoto networks of identical 2D phase oscillators with phase-lagged first-mode and higher-mode coupling.
We introduced the concept of cyclops states formed by two distinct, coherent clusters and a solitary oscillator reminiscent of the Cyclops's eye. These cyclops states represent a particular class of three-cluster generalized splay states \cite{berner2021generalized} with the solitary oscillator maintaining constant phase differences with the coherent clusters.
We reported a surprising finding that adding the second or third harmonic to the Kuramoto coupling makes the cyclops state global attractors in a wide range of couplings' repulsion. Beyond Kuramoto networks, we showed that the stabilization of cyclops states by the higher coupling harmonics is also robustly present in theta neurons with adaptive coupling. 

This paper extends our previous analysis to reveal higher-mode coupling-induced mechanisms for emerging new dynamical patterns termed breathing and switching cyclops states. Toward this goal, we derive analytical conditions on the existence and stability of cyclops states with 
constant inter-cluster phases in the presence of the second coupling harmonic. These conditions reveal two bifurcation scenarios for destabilizing the cyclops states. In the first scenario, the cyclops states with constant inter-cluster phases between its three clusters undergo an Andronov-Hopf bifurcation, preserving their intra-cluster formations but making the inter-cluster relative phase differences oscillate periodically.
Similarly to breathing three-cluster patterns introduced in \cite{brister2020three},
we call these breathing cyclops states. These states can evolve into roto-breathers with inter-cluster phase differences governed by mixed-mode, oscillatory-rotatory phase difference dynamics. In the second bifurcation scenario, the cyclops state with constant inter-cluster phases loses its structural stability but quickly reforms into a new cyclops state with a reshuffled configuration. This repetitive switching process yields switching cyclops states. These states are similar to blinking chimeras, also characterized by a death-birth process in which the coherent cluster dissolves and is quickly reborn in a new configuration \cite{goldschmidt2019blinking}.

We show that breathing, roto-breathing, and switching cyclops states are stable in a wide range
of the second harmonic coupling strength and phase lag parameter. Remarkably, breathing and roto-breathing cyclops states are dominant states, acting as the system's global attractors in a large interval of the second harmonic's phase lag parameter, corresponding to the overall repulsiveness of the combined first- and second-harmonic coupling. We also demonstrate that the cyclops states can robustly emerge in Kuramoto networks of non-identical oscillators. In \cite{munyayev2023cyclops}, we proved that 
the 2D Kuramoto model with the first and second-harmonic coupling is dynamically equivalent to a network of canonical theta-neurons with adaptive coupling. Therefore, we expect breathing and switching cyclops states to manifest strongly in theta-neuron networks, pointing to the broader applicability of our results.

The layout of this paper is as follows. In Sec.~II, we introduce the oscillator network model and state the problem under
consideration, and give formal definitions. 
In Sec.~III, we study the existence of cyclops states with constant inter-cluster phase differences, called stationary cyclops states. We derive an upper bound for the maximum number of stationary cyclops states with distinct inter-cluster phase differences. In Sec.~IV, we derive a four-dimensional (4D) system
that governs the dynamics of the inter-cluster phase differences. We study the stability of the fixed point of the 4D system, which corresponds to constant inter-cluster phase differences. We derive the conditions under which the fixed point undergoes an Andronov-Hopf bifurcation, giving rise to a breathing cyclops state. In Sec.~V, we analyze the variational equations for the transversal stability of stationary cyclops states that determines the stability of their coherent clusters. In Sec.~VI, we numerically study breathing and switching cyclops states emerging from stationary cyclops states via two distinct bifurcation routes. We demonstrate the emergence of more complex, hybrid dynamical patterns that combine the properties of both breathing and switching cyclops states. We also study the prevalence of different cyclops states and show that they robustly appear from large sets of randomly chosen initial conditions. In Sec.~VII,  we show the persistence of cyclops states against relatively large intrinsic frequency detuning.  Sec.~VIII contains concluding remarks and discussions. Appendix A
contains the derivation of the upper bound for the maximum number of stationary cyclops states.

\section{The model and problem statement}
We consider the Kuramoto-Sakaguchi network of 2D phase oscillators 
\begin{equation}
	\mu \ddot\theta_k+\dot\theta_k=\omega+\sum\limits_{n=1}^N \sum\limits_{q=1}^{2}\frac{\varepsilon_q}{N}
	{\sin\left[q\left(\theta_{n}-\theta_k\right)-\alpha_q\right]},	\label{eq:system_origin}
\end{equation}
where the $k$th oscillator's phase $\theta_k$ ranges from $-\pi$ to $\pi,$ and the second-order Kuramoto-Sakaguchi coupling \cite{sakaguchi2006instability} represents a pairwise interaction function $H(\theta_n-\theta_k)=\sum_{q=1}^{2}{\varepsilon_q}
	{\sin\left[q\left(\theta_{n}-\theta_k\right)-\alpha_q\right]}.$ The oscillators are assumed to be identical with frequency $\omega,$ inertia $m,$ and phase lag parameters $\alpha_1$ and $\alpha_2.$ We consider the phase lag $\alpha_1 \in (\pi/2, \pi),\;$
    that makes the first-harmonic coupling repulsive and fix $\varepsilon_1=1$ that corresponds to a strong  
 first-harmonic coupling. Throughout the paper, we also choose and fix a relatively strong inertia $\mu=1$ that is sufficient to make the dynamics of the 2D system qualitatively distinct from the 1D classical model and enable the emergence of breathing cluster dynamics \cite{belykh2016bistability}.
We will consider a broader range of $\alpha_2 \in (-\pi, \pi),$ so that the second harmonic may be pairwise attractive or repulsive. As a result, the overall combined coupling may be repulsive with $H'(0)<0$ or attractive with $H'(0)>0.$ The latter is possible when the second-harmonic coupling $\varepsilon_2$ is sufficiently strong to overcome the first-harmonic coupling contribution.

Phase coherence and cluster synchrony in the system~\eqref{eq:system_origin} can be characterized via the $l$th-order complex Kuramoto parameters \cite{daido1992order,skardal2011cluster}:
$$
R_l\!\left(t\right)=\frac{1}{N}\!\sum\limits_{k=1}^N\!{e^{i l\theta_{k}}}=r_l e^{i\psi_l},
$$
where $r_l$ and $\psi_l$ define the magnitude and the phase of
the $l$th moment Kuramoto order parameter $R_l(t)$, respectively. The first-order scalar parameter $r_1=|R_1|$ characterizes the degree of phase synchrony with $r_1=1$ corresponding to full phase
synchrony. Splay states or generalized splay states $\theta_k=\omega t+\varphi_k,$ $k=1,...,N$ with constant non-uniform relative phases $\varphi_k\in [-\pi,\pi]$ satisfy the condition  $r_1=0$ in the 2D Kuramoto model with the first-harmonic coupling ($\varepsilon_2=0$).  The second-order scalar parameter $r_2=|R_2|$ 
determines the degree of cluster synchrony. 
In the case of the first-harmonic coupling ($\varepsilon_2=0$), $r_2$ controls 
the stability of generalized splay states so that increasing $r_2$ enlarges their stability parameter regions \cite{berner2021generalized,munyayev2023cyclops}. It was shown in \cite{munyayev2023cyclops} that generalized splay states with a maximum $r_2$ are
(i) two-cluster symmetric splay states (for odd $N$) and (ii) three-cluster splay states with the relative phases (for even $N$):
\begin{equation}
	\begin{array}{l}
\varphi_1=\varphi_{2}=\ldots=\varphi_{M-1}=\gamma, \quad \varphi_M=0,\\
\varphi_{M+1}=\ldots=\varphi_{N}=-\gamma,
\end{array}
\label{cyclope}
\end{equation}
where $\gamma=\arccos\bigl(1\big/(1-N)\bigr),$  $M=(N+1)/2$, and the choice of the reference zero phase for $\varphi_M$ is arbitrary. We termed three-cluster splay states \eqref{cyclope} cyclops states.
Adding the second-harmonic coupling 
with $\varepsilon_2 \ne 0$ breaks their symmetry
in $\gamma$ and makes $r_1$ non-zero, albeit small. We demonstrated in \cite{munyayev2023cyclops} that the second- or higher-harmonic coupling can make these asymmetric patterns dominant states. In this paper, we generalize the definition of cyclops states \eqref{cyclope} for the system \eqref{eq:system_origin} with second-harmonic coupling and odd $N$ by relaxing the condition  $r_1=0.$ As a result, we refer to the following three-cluster state: 
\begin{equation}
	\begin{array}{l}
        \theta_1(t)=\theta_{2}(t)=\ldots=\theta_{M-1}(t)=x+\Omega t, \\
        \theta_M(t) = \Omega t, \\
        \theta_{M+1}(t)=\theta_{M+2}(t)=\ldots=\theta_{N}(t)=y+\Omega t
    \end{array}
    \label{stat_cyclope_nonsym}
\end{equation}
as to a {\it stationary cyclops state} in which two equal clusters of $M-1$ oscillators rotate with the common frequency $\Omega,$ preserving the stationary phase differences $x=\gamma_1,$ and $y=\gamma_2$ 
with the $M$th solitary oscillator ($x \ne y$). The common rotational frequency $\Omega$ can be explicitly calculated from \eqref{eq:system_origin} (see the next section). Due to the system's global coupling symmetry and equal cluster sizes $M-1,$ the existence of a stationary cyclops state with inter-cluster phase differences $x=\gamma_1$ and $y=\gamma_2$ implies the existence of its counterpart with $x=\gamma_2$ and $y=\gamma_1$. Thus, cyclops states exist in symmetrical pairs. 
\begin{figure}[h!]\center
\includegraphics[width=0.60\columnwidth]{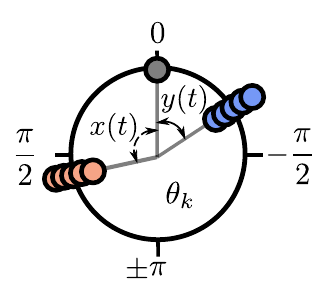}
	\caption{Snapshot of a breathing cyclops state in network~\eqref{eq:system_origin} of $11$ oscillators. Periodically oscillating $x(t)$ and $y(t)$ govern the phase difference between the synchronous clusters (blue and pink circles) and the solitary oscillator (gray circle). The solitary oscillator's phase is chosen at $\theta_M=0$ as a reference. Parameters are $\alpha_1=1.7$, $\varepsilon_2=0.08,$ $\alpha_2=-0.3$.}
\label{fig0}
\end{figure}
In the following, we will analyze the existence and stability of stationary cyclops states in the system \eqref{eq:system_origin} with odd $N.$ We will describe two main scenarios for destabilizing a stationary cyclops state that yield: \\
(i) {\it a breathing cyclops state} with periodically oscillating $x(t)$, $y(t)$ so that the inter-cluster phase differences are bounded as $|x(t)|< \pi$, $|y(t)|< \pi$ to produce no phase slips (Fig.~\ref{fig0}); \\
(ii) {\it a switching cyclops state}, representing a repetitive death-birth process in which the clusters disintegrate to form a new cyclops state with a new reshuffled cluster configuration and a new solitary node.

We will also study how the breathing and switching cyclops states can (i) further evolve into more complex dynamical patterns, including hybrid switching-breathing states, (ii) become globally stable, and (iii) persist against intrinsic frequency detuning.

\section{Possible constant inter-cluster phase differences}\label{possible}
We seek to determine permissible stationary cyclops states as a function of the system's parameters. To determine the constant phase differences $x,$ $y,$ and the rotational frequency $\Omega,$ we substitute the stationary cyclops state solution \eqref{stat_cyclope_nonsym} into \eqref{eq:system_origin} and obtain the  system of nonlinear transcendental equations:
\begin{equation}
\begin{aligned}
    \omega - \Omega - \sum_{q=1}^{2} \frac{\varepsilon_q}{N} \Bigl[ \sin(qx+\alpha_q) + \frac{N\!-\!1}{2} \left( \sin\alpha_q\right.\Bigr.\\
    \Bigl.\left.+ \sin(q(x-y)\!+\!\alpha_q) \right)\Bigr]=0,\\
    \omega - \Omega - \sum_{q=1}^{2} \frac{\varepsilon_q}{N} \Bigl[ \sin \alpha_q -\frac{N\!-\!1}{2} \left( \sin(qx\!-\!\alpha_q)\Bigr.\right.\\
    \Bigl.\left.+ \sin(qy\!-\!\alpha_q) \right)\Bigr]=0,\\
    \omega - \Omega - \sum_{q=1}^{2} \frac{\varepsilon_q}{N} \Bigl[ \sin(qy+\alpha_q) + \frac{N\!-\!1}{2} \left( \sin\alpha_q\Bigr.\right.\\
    \Bigl.\left.+ \sin(q(y-x)\!+\!\alpha_q) \right)\Bigr]=0.\\
\end{aligned}
\label{eq:cyclope_find_nonlin_full}
\end{equation}
We subtract the second equation from the first and third equations of \eqref{eq:cyclope_find_nonlin_full} to eliminate $\Omega$ and obtain the system of two equations for finding the unknown constants $x$ and~$y$:
\begin{equation}
    \begin{aligned}
        &\frac{N-3}{2}\sum_{q=1}^{2}{\varepsilon_q}
	{\sin\alpha_q}+\sum_{q=1}^{2}{\varepsilon_q}
	{\sin(qx+\alpha_q)}\\
        &+\frac{N-1}{2}\Big(\sum_{q=1}^{2}{\varepsilon_q}
	{\sin(qx-\alpha_q)}+\sum_{q=1}^{2}{\varepsilon_q}
	{\sin(qy-\alpha_q)}\\
    &-\sum_{q=1}^{2}{\varepsilon_q}
	{\sin(q(y-x)-\alpha_q)}\Big)=0,\\
    &\frac{N-3}{2}\sum_{q=1}^{2}{\varepsilon_q}
    	{\sin\alpha_q}+\sum_{q=1}^{2}{\varepsilon_q}
    	{\sin(qy+\alpha_q)}\\
            &+\frac{N-1}{2}\Big(\sum_{q=1}^{2}{\varepsilon_q}
    	{\sin(qx-\alpha_q)}+\sum_{q=1}^{2}{\varepsilon_q}
    	{\sin(qy-\alpha_q)}\\
        &-\sum_{q=1}^{2}{\varepsilon_q}
    	{\sin(q(x-y)-\alpha_q)}\Big)=0.
    \end{aligned}
    \label{eq:cyclope_find_nonlin}
\end{equation}
In turn, we find the rotational frequency $\Omega$ from the second equation of \eqref{eq:cyclope_find_nonlin_full}:
\begin{equation}
    \begin{aligned}
        &\Omega = \omega - \frac{1}{N} \sum_{q=1}^{2} \varepsilon_q \sin \alpha_q \\
        &+\frac{N\!-\!1}{2N} \left[ \sum_{q=1}^{2} \varepsilon_q \sin(qx\!-\!\alpha_q) + \sum_{q=1}^{2} \varepsilon_q \sin(qy\!-\!\alpha_q) \right]
    \end{aligned}
    \label{eq:omega}
\end{equation}
with $x$ and $y$ calculated from  \eqref{eq:cyclope_find_nonlin}.

Due to the complexity of system~\eqref{eq:cyclope_find_nonlin}, its solution for $x$ and $y$ cannot be found in closed form.
Yet, we derive an upper bound for the maximum number of stationary cyclops states with distinct $x$ and $y.$ To do so, 
we transform the real-valued system \eqref{eq:cyclope_find_nonlin} into a system of complex polynomial equations and apply the Bernshtein theorem \cite{bernshtein1975number}, a practical tool in algebra that bounds the number of non-zero complex solutions by the mixed volume of their Newton polytopes. The details of this analysis are quite technical and are delegated to Appendix A. This analysis shows that the complex form of system ~\eqref{eq:cyclope_find_nonlin} may have up to $17$ possible solutions (including some non-physical) corresponding up to $16$ stationary cyclops states with distinct ordered pairs of constant phase differences $x,y.$ As stationary cyclops states exist in pairs, there may be at most $8$ combinations of $x,y$ (up to the cluster permutation  $x \longleftrightarrow y$). It is worth noticing that there is a continuum of stationary cyclops states with a given pair $(x,y)$ due to an arbitrary choice of the reference solitary state's phase $\theta_M.$
\begin{figure}[b!]\center
\includegraphics[width=0.95\columnwidth]{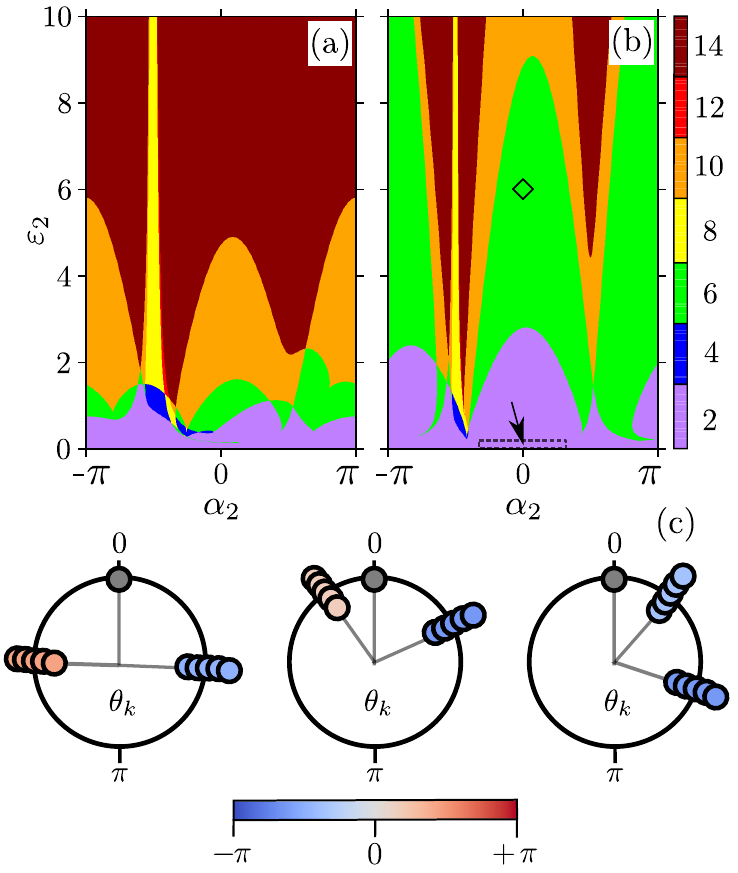}
	\caption{(a, b). The color shows the number of distinct stationary cyclops states in the network~\eqref{eq:system_origin} as a function of the second-harmonic coupling and phase lag parameters $(\alpha_2, \varepsilon_2).$ Other parameters are
    (a) $N=5$, $\alpha_1=2.0$ and (b) $N=11$, $\alpha_1=1.7$. The number of cyclops is calculated by numerically finding solutions of system~\eqref{eq:cyclope_find_nonlin}. 
    The arrow points to the dashed area corresponding to the stability diagram of Fig.~\ref{fig1}. (c). Snapshots of  
    three distinct stationary cyclops states (up to permutation of clusters $x \longleftrightarrow y$) for the parameter set $N=11$, $\alpha_1=1.7$, $\alpha_2=0.0$, $\varepsilon_2=6.0$ corresponding to the open diamond in the green area in panel b. The oscillator coloring corresponds to the inter-cluster differences $x$ and $y$ according to positive and negative values depicted from the horizontal color bar.}
\label{fig:number_cyclops}
\end{figure}

Figure~\ref{fig:number_cyclops} displays the number of different stationary cyclops states calculated by solving the complex polynomial equation \eqref{eq:cyclope_find_lin} using the NSolve function of Wolfram Mathematica. Note that this number equals two for small values of the second-harmonic amplitude $\varepsilon_2.$ This pair of stationary cyclops states with $x=\gamma_1$ and $y=\gamma_2$ ($x=\gamma_2$ and $y=\gamma_1$) emerges continuously from the symmetrical cyclops state \eqref{cyclope} that exists in the system \eqref{eq:system_origin} in the absence of the second-harmonic coupling ($\varepsilon_2=0$). As Figs.~\ref{fig:number_cyclops}a,b indicate, increasing $\varepsilon_2$ increases the number of co-existing stationary cyclops states and induces richer dynamics.

\section{Stability of the inter-cluster phase differences}
We seek to derive the conditions for the stability of the constant inter-cluster phase differences to small perturbations of $x$ and $y.$ The dynamics of the inter-cluster phase differences are governed by the system: 
\begin{equation}
\begin{array}{lcl}
    \mu\ddot{x}+\dot{x}&=&\sum \limits_{q=1}^{2} \displaystyle\frac{\varepsilon_q}{N}[\sin \alpha_q - \sin(qx+\alpha_q)\\
	&&-\displaystyle\frac{N\!-\!1}{2} ( \sin(qx\!-\!\alpha_q) + \sin(qy\!-\!\alpha_q)\\
    &&+\sin\alpha_q+\sin(q(x-y)\!+\!\alpha_q) )],\\
	\mu\ddot{y}+\dot{y}&=&\sum \limits_{q=1}^{2}\displaystyle\frac{\varepsilon_q}{N} [\sin \alpha_q - \sin(qy+\alpha_q)\\
	&&-\displaystyle\frac{N\!-\!1}{2} ( \sin(qx\!-\!\alpha_q) + \sin(qy\!-\!\alpha_q)\\
    &&+\sin\alpha_q + \sin(q(y-x)\!+\!\alpha_q) )].
\end{array}
\label{eq:breatherXY}
\end{equation}
The 4D dynamical system \eqref{eq:breatherXY} may be viewed loosely as a system of two nonlinearly coupled driven pendulum-like equations with the terms $\sin \alpha_q$ representing constant torques and the sine terms with $x$ and $y$ corresponding to pendulum-like nonlinearities and coupling. The presence of the second-harmonic coupling prevents transforming the system \eqref{eq:breatherXY} into a more explicit system of two coupled pendula as it was achieved for a three-cluster state in \cite{brister2020three}. However, the pendulum-like structure of the 4D system \eqref{eq:breatherXY} points to the possible existence of nontrivial dynamics related to oscillating and even chaotically evolving intercluster phase differences $x(t)$ and $y(t).$

Fixed points of system \eqref{eq:breatherXY} correspond to constant inter-cluster phase differences $x,y$ calculated from \eqref{eq:cyclope_find_nonlin}. We aim to study 
the local stability of the fixed points and derive bifurcation conditions that induce oscillating phase differences $x(t),y(t).$ Toward this goal, we consider small deviations $\delta x(t)$ and $\delta y(t)$ from a fixed point $x=\gamma_1$, $y=\gamma_2$ corresponding to a stationary cyclops state. So,  $x(t)=\gamma_1 + \delta x(t)$, $y(t)=\gamma_2 + \delta y(t)$. We linearize the system~\eqref{eq:breatherXY} in the vicinity of the fixed point  state and obtain the following equations that govern the evolution of small deviations $\delta x(t)$ and $\delta y(t)$:
\begin{widetext}
	\begin{equation}
		\begin{aligned}
			\mu\delta\ddot{x}+\delta\dot{x}=-\sum_{q=1}^{2} \frac{\varepsilon_q q}{N} \Bigl[\cos(q\gamma_1+\alpha_q) \delta x
			+ \frac{N\!-\!1}{2} \left( \cos(q\gamma_1\!-\!\alpha_q) \delta x + \cos(q\gamma_2\!-\!\alpha_q) \delta y
			+ \cos(q \sigma\!+\!\alpha_q) (\delta x - \delta y) \right)\Bigr],\\
			\mu\delta\ddot{y}+\delta\dot{y}=-\sum_{q=1}^{2} \frac{\varepsilon_q q}{N} \Bigl[\cos(q\gamma_2+\alpha_q) \delta y
			+ \frac{N\!-\!1}{2} \left( \cos(q\gamma_1\!-\!\alpha_q) \delta x +\cos(q\gamma_2\!-\!\alpha_q) \delta y
			+ \cos(q \sigma\!-\!\alpha_q) (\delta y - \delta x) \right)\Bigr],\\
		\end{aligned}\label{variational}
	\end{equation}
 \end{widetext}
 where $\sigma=\gamma_1-\gamma_2.$
 
 Following the standard stability approach, we seek solutions $\delta x(t)=A_1 e^{\lambda t},$ $\delta y(t)=A_2 e^{\lambda t}$ and derive a system of two characteristic equations for finding constants $\lambda,$ $A_1$ and $A_2$:
 	\begin{equation}
		\begin{aligned}
			(\mu\lambda^2+\lambda)A_1=-(p_{11} A_1 + p_{12} A_2),\\
			(\mu\lambda^2+\lambda)A_2=-(p_{21} A_1 + p_{22} A_2),\\
		\end{aligned}\label{characteristics}
	\end{equation}
	where
%\begin{equation}
%\begin{aligned}
%    p_{11}\!=\!\sum_{q=1}^{2}\!\frac{\varepsilon_q q}{N} \Bigl[\cos(q\gamma_1+\alpha_q) + \frac{N\!-\!1}{2} \left( \cos(q\gamma_1\!-\!\alpha_q)\right.\Bigr.\\
%	\Bigl.\left.+ \cos(q\sigma\!+\!\alpha_q) \right)\Bigr],\\
%	p_{12}=\sum_{q=1}^{2} \frac{\varepsilon_q q}{N} \Bigl[\frac{N\!-\!1}{2} \left(\cos(q\gamma_2\!-\!\alpha_q) - \cos(q \sigma\!+\!\alpha_q) \right)\Bigr],\\
%	p_{21}=\sum_{q=1}^{2} \frac{\varepsilon_q q}{N} \Bigl[\frac{N\!-\!1}{2} \left( \cos(q\gamma_1\!-\!\alpha_q) - \cos(-q \sigma\!+\!\alpha_q) \right)\Bigr],\\
%	p_{22}=\sum_{q=1}^{2} \frac{\varepsilon_q q}{N} \Bigl[\cos(q\gamma_2+\alpha_q) + \frac{N\!-\!1}{2} \left(\cos(q\gamma_2\!-\!\alpha_q)\right.\Bigr. \\
%    \Bigl.\left.+ \cos(-q \sigma\!+\!\alpha_q) \right)\Bigr].\\
%\end{aligned}
%\label{ps}
%\end{equation}
\begin{equation}
\begin{aligned}
    p_{11}\!=\!\sum_{q=1}^{2}\!\frac{\varepsilon_q q}{N}\Bigl[\frac{N\!-\!1}{2} \big(\!\cos(q\gamma_1\!-\!\alpha_q)\!+\!\cos(q\sigma\!+\!\alpha_q)\big)\\+\cos(q\gamma_1+\alpha_q)\Bigr],\\
	p_{12}\!=\!\sum_{q=1}^{2}\!\frac{\varepsilon_q q}{N}\Bigl[\frac{N\!-\!1}{2} \left(\cos(q\gamma_2\!-\!\alpha_q)\!-\!\cos(q\sigma\!+\!\alpha_q)\right)\Bigr],\\
	p_{21}\!=\!\sum_{q=1}^{2}\!\frac{\varepsilon_q q}{N}\Bigl[\frac{N\!-\!1}{2} \left(\cos(q\gamma_1\!-\!\alpha_q)\!-\!\cos(q \sigma\!-\!\alpha_q)\right)\Bigr],\\
	p_{22}\!=\!\sum_{q=1}^{2}\!\frac{\varepsilon_q q}{N}\Bigl[\frac{N\!-\!1}{2} \big(\cos(q\gamma_2\!-\!\alpha_q)\!+\!\cos(q \sigma\!-\!\alpha_q)\big)\\+\cos(q\gamma_2+\alpha_q)\Bigr].\\
\end{aligned}
\label{ps}
\end{equation}
Solving the characteristic system~\eqref{characteristics} of two coupled quadratic equations to explicitly find $\lambda$ is out of reach. Instead, we introduce the new variable $\Lambda=\mu\lambda^2+\lambda$ and turn
 the system \eqref{characteristics} into the system of linear equations
	\begin{equation}
		\boldsymbol{P} (A_1, A_2)^{T} = \Lambda (A_1, A_2)^{T},
	\end{equation}
	where
	\begin{equation}
		\boldsymbol{P}=
		\begin{pmatrix}
			-p_{11} & -p_{12}\\
			-p_{21} & -p_{22}
		\end{pmatrix}.\label{eigen}
	\end{equation}
Therefore, the stability of \eqref{variational} can be assessed from \eqref{eigen} in terms of its eigenvalues $\Lambda.$ To do so,
we aim to determine the boundary of the stability region that is determined by $\lambda = i \text{ Im}\lambda$ and corresponds to 
an Andronov-Hopf bifurcation of the fixed point that induces oscillating phase differences $x(t)$ and $y(t).$ Therefore, we can set $\text{Re} \Lambda + i \text{ Im} \Lambda = -\mu(\text{Im}\lambda)^2 + i\text{ Im}\lambda$ so that the real part equality $\text{Re}\Lambda+\mu(\text{Im}\Lambda)^2=0$ defines
the stability boundary  $\lambda = i \text{Im}\lambda.$ We pick the mass $\mu=0$ as a test number that makes $\Lambda = \lambda$ and yields the inequality $\text{Re} \Lambda <0$ for the fixed point stability. Extending this argument to non-zero $\mu,$ we can conclude that the inequality 
$\text{Re}\Lambda+\mu(\text{Im}\Lambda)^2<0$ makes $\text{Re} \lambda <0$ and determines the fixed point stability. Thus, we arrive at the following assertion.\\
{\bf Statement 1.} [Internal stability of stationary cyclops states].\\
{\it 1. Constant inter-cluster phase differences $x\!=\!\gamma_1$ and~$y\!=\!\gamma_2$~of the stationary cyclops state \eqref{stat_cyclope_nonsym}
are\,locally\,stable\,iff
\begin{equation}
\begin{aligned}
    \text{\rm Re} \Lambda_{1,2}+\mu(\text{\rm Im}\Lambda_{1,2})^2<0,\\
    \Lambda_{1,2}=\frac{\text{\rm Tr}\boldsymbol{P}}{2}\pm\frac{\sqrt{(\text{\rm Tr}\boldsymbol{P})^2-4 \text{\rm det} \boldsymbol{P}}}{2},
\end{aligned}
\label{statement1}
\end{equation}
where $\text{\rm Tr}\boldsymbol{P}$ and $\text{\rm det}\boldsymbol{P}$ are, respectively, the trace and determinant of matrix $\boldsymbol{P}$~\eqref{eigen} whose coefficients are defined in \eqref{ps}.\\
  2. The stability boundary 
  \begin{equation}
\text{\rm Re}\Lambda_{1,2}+\mu(\text{\rm Im}\Lambda_{1,2})^2=0 \label{hopf}
	\end{equation}
corresponds to an Andronov-Hopf bifurcation that destabilizes the stationary cyclops state, turning it into a breathing cyclops state with oscillating inter-cluster phase differences $x(t)$ and $y(t).$}

\begin{figure}[h!]
	\includegraphics[width=0.80\columnwidth]{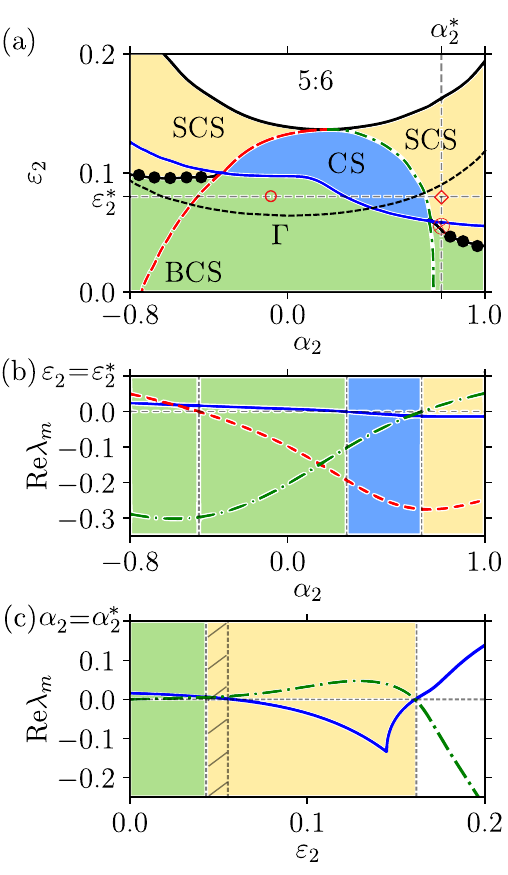}
	\caption{(a). The stability diagram for cyclops states. Regions of stable stationary cyclops states (CS) are shown in blue, switching cyclops states (SC) in yellow, breathing cyclops states (BCS) in green, and two-cluster regimes ($5:6$) in white. Analytical boundaries: the blue solid line corresponds to the stability boundary \eqref{hopf}, the red dashed line to $\text{Re}\lambda^{tran}_{1}=0$, and the green dash-dotted curve to $\text{Re}\lambda^{tran}_{1}=0$. Two numerical curves marked by the solid circles separate the stability regions of the switching and breathing stationary cyclops states.
  The black dotted line $\Gamma$ corresponds to $H'(0)=0.$ Values above the curve make the coupling attractive and full synchronization locally stable.
Stationary cyclops states are found as a solution of  system~\eqref{eq:cyclope_find_nonlin} and used as initial conditions. The round, diamond, and circled times correspond to the parameters used in 
Figs.~\ref{fig2},~\ref{fig3},~\ref{fig4}.  (b). The real part of the eigenvalues, associated with the stationary cyclops state, 
that determine the (internal) stability of the inter-cluster phase differences
(blue solid line) and transversal (external) stability of the first (red dashed line) and second cluster (green dash-dotted line) for fixed $\varepsilon_2=\varepsilon^*_2$ and varying $\alpha_2$ (along the white dashed horizontal line in panel a). The background color indicates the type of the emerged cyclops states as in panel a. (c). The diagram is similar to panel b, but for fixed $\alpha_2=\alpha^*_2$ and varying $\varepsilon_2$ (along the black-white dashed vertical line panel a). The shaded area indicates the bistability of switching and breathing cyclops states. Parameters: $N=11$, $\varepsilon_1=1.0$, $\alpha_1=1.7$, $\varepsilon^*_2=0.08$, $\alpha^*_2=0.78$.}
	\label{fig1}
\end{figure}

The stability of constant phase differences $x$ and $y$ defined via \eqref{statement1} can be interpreted as the internal (longitudinal) stability of the stationary cyclops state within the invariant three-cluster manifold $D$ determined by \eqref{stat_cyclope_nonsym} with arbitrary, possibly time-varying $x(t)$ and $y(t).$ The stability boundary
\eqref{hopf} depicted by the blue solid curve in Fig.~\ref{fig1}a corresponds to emerging instability of constant phase differences $x$ and $y$ that preserves the three-cluster cyclops formation on the invariant three-cluster manifold $D.$

Having studied the stability of the constant inter-cluster phase differences, we proceed with the stability analysis of the synchronous clusters, each composed of 
$M-1$ oscillators. These conditions, paired with the condition \eqref{statement1}, shall indicate what stationary cyclops states can stably emerge in the
network.

\section{Stability of synchronous clusters}
We aim to derive the conditions for transversal stability of the stationary cyclops state     \eqref{stat_cyclope_nonsym} that amounts to the stability of the two synchronous clusters composing the stationary cyclops state. We introduce small deviations from the oscillators' phases $\theta_n \longrightarrow \theta_n + \delta \theta_n,$ composing the first cluster for  $n=1,...,M-1$ and the second cluster for $n=M+1,...,N.$ 
To study the local stability of each synchronous cluster,
we consider the difference variables 
\begin{equation}
	\xi_n=\delta\theta_{n+1}-\delta\theta_{n},\quad n=1,\dots, M-2,\label{cluster1}
\end{equation}
\begin{equation}
	\zeta_{n}=\delta\theta_{n+1}-\delta\theta_{n},\quad n=M+1,\dots, N-1 \label{cluster2}
\end{equation}
that describe the phase difference dynamics within the first and second clusters, respectively. Therefore, from \eqref{eq:system_origin}, \eqref{stat_cyclope_nonsym}, and \eqref{cluster1}, \eqref{cluster2}, we obtain two uncoupled variational equations with time-invariant coefficients. Each of the equations determines the local stability of the corresponding cluster within the cyclops state \eqref{stat_cyclope_nonsym}:
\begin{equation}
	\begin{aligned}
		\mu\ddot\xi_n + \dot\xi_n + \frac{1}{N}\sum_{q=1}^{2}{\varepsilon_q}q\Big[
		{\cos(q \gamma_1+\alpha_q)}
		+\\ \frac{N-1}{2}(
		{\cos\alpha_q}+
		{\cos(q \sigma+\alpha_q)}) \Big] \xi_n = 0,
	\end{aligned}
	\label{eq:system_variation_2_2}
	\end{equation}
	where $n=1,2,\dots,M-2$, and
	\begin{equation}
	\begin{aligned}
		\mu\ddot\zeta_n + \dot\zeta_n + \frac{1}{N}\sum_{q=1}^{2}{\varepsilon_q}q\Big[
		{\cos(q \gamma_2+\alpha_q)}
		+\\ \frac{N-1}{2}({\cos\alpha_q}+
		{\cos(q \sigma-\alpha_q)}) \Big] \zeta_n = 0,
	\end{aligned}
	\label{eq:system_variation_2_3}
	\end{equation}
	where $n=M+1,\dots, N-1$.
 	The variational equations \eqref{eq:system_variation_2_2} and \eqref{eq:system_variation_2_3} are stable iff
   the time-invariant coefficients of the terms $\xi_n$ and $\zeta_n$ are positive. Therefore, we can formulate the stability conditions in the following assertion. \\
   {\bf Statement 2.} [Transversal stability of stationary cyclops states].
{\it Clusters of oscillators composing the stationary cyclops state \eqref{stat_cyclope_nonsym} are locally stable iff:
\begin{equation}
	\begin{aligned}
        \sum_{q=1}^{2}{\varepsilon_q} {q\cos(q \gamma_1+\alpha_q)}+\frac{N-1}{2}\Bigl(\sum_{q=1}^{2}{\varepsilon_q} {q\cos\alpha_q}+\Bigr.\\
        \Bigl.\sum_{q=1}^{2}{\varepsilon_q} {q\cos(q \sigma+\alpha_q)}\Bigr)>0,\\
        \sum_{q=1}^{2}{\varepsilon_q} {q\cos(q \gamma_2+\alpha_q)}+\frac{N-1}{2}\Bigl(\sum_{q=1}^{2}{\varepsilon_q} {q\cos\alpha_q}+\Bigr.\\
        \Bigl.\sum_{q=1}^{2}{\varepsilon_q} {q\cos(q \sigma-\alpha_q)}\Bigr)>0,
\end{aligned}\label{statement2}
\end{equation}
where the right-hand sides of the inequalities \eqref{statement2} are the coefficients of the variational equations \eqref{eq:system_variation_2_2} and \eqref{eq:system_variation_2_3}.}

It is also straightforward to show that the stationary cyclops state is always stable to the shift of all phases by a constant value $\delta\theta_k=\delta\theta$ ($k=1,\dots,N$). 

It is worth noticing that the eigenvalues $\lambda^{tran}_{1,2}$ associated with the variational equations \eqref{eq:system_variation_2_2} and \eqref{eq:system_variation_2_3} have multiplicity $M-2$.
Thus, the eigenvalues $\lambda^{tran}_{1}$ and $\lambda^{tran}_{2}$ define the transversal stability of the first ($n=1,2,\dots,M-2$) and second ($n=M+1,\dots,N$) clusters, respectively. 
 Figure~\ref{fig1}a displays their stability boundaries defined by the conditions \eqref{statement2} with the left-hand sides set to 0 to  correspond to $\text{Re}\lambda^{tran}_{1}=0$ (the red dashed line) and 
$\text{Re}\lambda^{tran}_{2}=0$ (the green dash-dotted line).
To highlight the constructive role of the second-harmonic coupling with $\varepsilon_2 \ne 0$, we chose the parameter values that yield unstable stationary cyclops in the network with only first-harmonic coupling with $\varepsilon_2 =0$ (see Fig.~\ref{fig1}).

As Fig.~\ref{fig1}a indicates, crossing the stability boundary \eqref{hopf} (the lower border of the region CS) induces breathing cyclops states in the region BCS (green) in accordance with Statement~1.  In turn, crossing the upper border of the region CS, composed of the transveral stability boundaries $\text{Re}\lambda^{tran}_{1}=0$ (the red dashed line) and $\text{Re}\lambda^{tran}_{2}=0$ (the green dash-dotted line) can yield either switching cyclops states in the region SCS (yellow) or asymmetrical, two-cluster states with five- and six-oscillator synchronous clusters (white region $5:6$). In the following, we will primarily focus on the properties of emerging breathing and switching cyclops states.

\begin{figure}[h!]\center
\includegraphics[width=0.90\columnwidth]{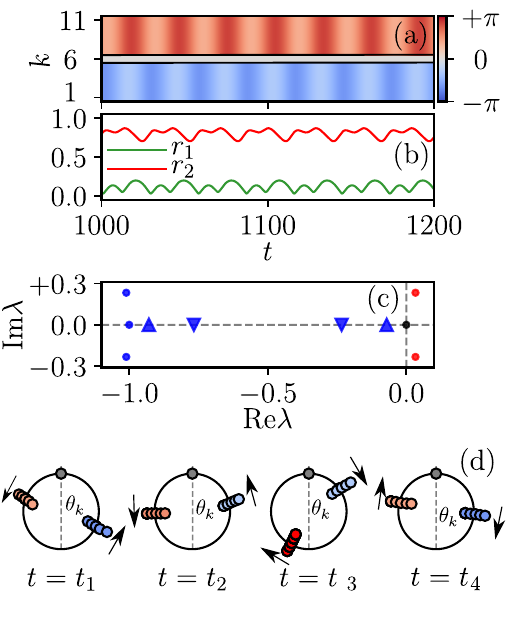}
	\caption{Breathing cyclops state. (a). The colors depict the phase differences $\theta_k(t) - \theta_{6}(t).$ The gray strip indicates the reference solitary oscillator. (b). The corresponding values of $r_1$ and $r_2$. (c). The eigenvalues associated with the destabilized stationary cyclops state. Some eigenvalues are repeated.  The round (triangular) labels correspond to the internal (transversal) stability. Note a pair of complex eigenvalues with a positive real part (red) that emerged due to an Andronov-Hopf bifurcation and yielded periodic oscillations of inter-cluster differences. (d). Phase distributions $\theta_k$ at several time instants. The arrows indicate the direction of periodic phase clusters' oscillations (see Supplementary Movie 1 demonstrating this breathing cyclops state). The oscillators' coloring represents their relative phase difference with the solitary oscillator as in Fig.~\ref{fig:number_cyclops}c.  Parameters $N=11$, $\mu=1.0$, $\varepsilon_1=1.0$, $\alpha_1=1.7$, $\varepsilon_2=0.08$, $\alpha_2=-0.1$ correspond to the open circle label in Fig.~\ref{fig1}a.}
\label{fig2}
\end{figure}

\begin{figure*}[t!]
	\includegraphics[width=0.85\textwidth]{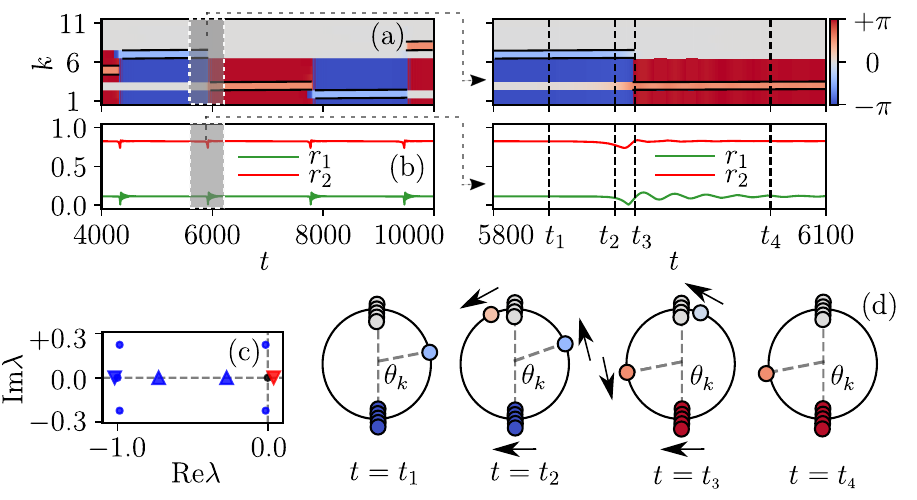}
	\caption{Switching cyclops state. (a). The colors depict the phase differences $\theta_k(t) - \theta_{6}(t).$ The strips with solid black borders indicate the reference solitary oscillator during the lifetime of a cyclops state configuration (the first stage). Note that clusters disintegrate to form a new cyclops state with a different solitary oscillator (the second stage). (b). The corresponding values of $r_1$ and $r_2$. The gray fragments correspond to the zoomed-in insets (right panels). (c). The eigenvalues associated with the destabilized stationary cyclops state. Some eigenvalues are repeated. The round (triangular) labels correspond to the internal (transversal) stability. Note a positive real eigenvalue (red) corresponding to the loss of the transversal stability of the stationary cyclops state due to Statement~2. (d). Phase distributions $\theta_k$ corresponding to a death-birth process in which a cyclops state existing at $t=t_1$ disintegrates to form a new cyclops state at $t=t_4$ (see Supplementary Movie 2 for the details of this dynamical evolution). Parameters $N=11$, $\mu=1.0$, $\varepsilon_1=1.0$, $\alpha_1=1.7$, $\varepsilon_2=0.08$, $\alpha_2=0.78$ correspond to the diamond label in Fig.~\ref{fig1}a.}
	\label{fig3}
\end{figure*}

\begin{figure*}[t!]\center
	\includegraphics[width=0.85\textwidth]{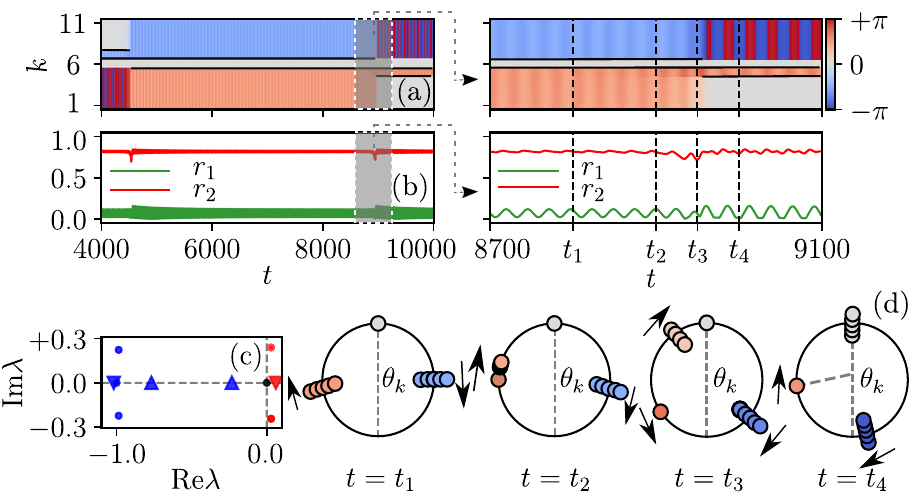}
	\caption{Switching-breathing cyclops state. The notations are as in Fig.~\ref{fig3}. One cluster of the breathing cyclops state (depicted in orange in panel a) eventually disintegrates, forming a reshuffled synchronous cluster and a new solitary oscillator. Note the weak internal and transversal instability of the destabilized stationary cyclops state due to the three eigenvalues with small positive real parts (red circles and nabla in panel c). Supplementary Movie 3 animates the sequence given in panel d. Parameters $N=11$, $\mu=1.0$, $\varepsilon_1=1.0$, $\alpha_1=1.7$, $\varepsilon_2=0.0578$, $\alpha_2=0.78$ correspond to the circled times label in Fig.~\ref{fig1}a.}
 \label{fig4}
\end{figure*}

\section{Emerging breathing and switching cyclops states}
Figure \ref{fig1} confirmed the two main bifurcation scenarios for destroying the stationary cyclops states
and generating breathing and switching cyclops states
described by Statements 1 and 2. In the first scenario, complex conjugate eigenvalues 
$\lambda_{1,2}$, that determine the stability of constant inter-cluster phase differences $x$ and $y$ via \eqref{characteristics},
become purely imaginary and induce oscillating  $x(t)$ and $y(t)$ (Fig.~\ref{fig2}c). As a result,
the stationary cyclops state becomes internally unstable; however, the stability of the clusters preserves and guarantees the emergence of a breathing cyclops state (see Fig.~\ref{fig2}a,d and Supplementary Movie 1 for the animation of the breathing cyclops state dynamics). Periodic oscillations of the first two order parameters $r_1$ and $r_2$ depicted in Fig.~\ref{fig2}b are a 
signature of such a breathing cyclops state.
As the distance from the stability boundary of the $CS$ region (solid blue line in Fig.~\ref{fig1}a) increases when changing the second-harmonic coupling strength  $\varepsilon_2$ and phase lag $\alpha_2,$ the amplitudes of inter-cluster difference oscillations $x(t)$, $y(t)$, and order parameters $r_1(t)$, $r_2(t)$ increase. It is worth noticing that for the parameters $\alpha_1$ and $\varepsilon_1$ used in Fig.~\ref{fig1}a, the breathing cyclops state is also stable in the absence of the second-harmonic coupling ($\varepsilon_2=0$). 

In the second bifurcation scenario determined via Statement~2, the stationary cyclops state loses its transversal stability when one of the eigenvalues 
$\lambda^{tran}_{1,2}$ becomes positive (Fig.~\ref{fig3}c). Note that the real parts of the other eigenvalues controlling the internal stability of the inter-cluster differences remain negative, thereby preserving the stable component of the saddle dynamics. While the transversal instability of the cyclops state may lead to its complete destruction, it induces a switching cyclops state (Fig.~\ref{fig3}a,d) when the transversal instability is weak (note the slightly positive eigenvalue, depicted by the red nabla in Fig.~\ref{fig3}c). This non-stationary cyclops state represents a two-stage repetitive process. During the first relatively long 
stage, the inter-cluster differences
$x$ and $y$ practically do not change, and the synchronous clusters preserve their formation, i.e., the dynamical pattern is similar to a stationary cyclops state (Fig.~\ref{fig3}). During the second short stage, one cluster reshuffles so that one node leaves the unstable cluster to become a new solitary oscillator, whereas the remaining oscillators from the cluster merge with the old solitary node. Figure~\ref{fig3}d and Supplementary Movie 2 illustrate this process. Accordingly, during the first stage, the magnitudes of the order parameters $r_1$ and $r_2$ are practically constant. They undergo an abrupt change during the second stage to return to a constant value (Fig.~\ref{fig3}b). The duration of the first stage, and, hence, the period of oscillations in $r_1$ and $r_2$, gradually decreases as the instability develops when the transversal 
eigenvalue becomes more positive under the control parameters change. Eventually, the switching cyclops state turns into a chaotically switching dynamical pattern.

We also observe a hybrid of the switching and breathing cyclops states (Fig.~\ref{fig4}b). This hybrid state emerges when, in addition to the external instability of one cluster, there is an internal instability of the inter-cluster phase differences $x$ and $y$ (Fig.~\ref{fig1}). In terms of the eigenvalue spectrum, this amounts to the presence of a pair of complex conjugate eigenvalues $\lambda_{1,2}$ (corresponding to the internal instability) and one real eigenvalue $\lambda^{tran}_{1}$ lying to the right from the imaginary axis (Fig.~\ref{fig4}c). 
We term this hybrid a {\it switching-breathing cyclops state}, which is effectively a switching cyclops state, which, during its first stage, has oscillating inter-cluster phase differences $x(t)$ and $y(t)$. Accordingly, the order parameter amplitudes $r_1$ and $r_2$ are time-periodic functions (Fig.~\ref{fig4}b). Figure~\ref{fig4}d and Supplementary Movie 3 detail the dynamical evolution of the switching-breathing cyclops state.

\begin{figure}[h!]\center
\includegraphics[width=0.90\columnwidth]{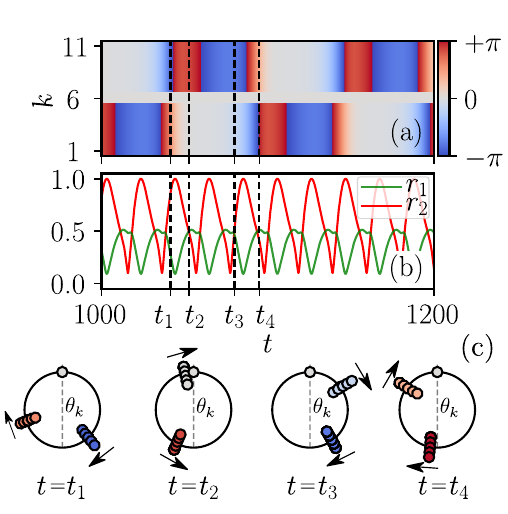}
	\caption{Rotobreathing cyclops state.  The notations are as in Fig.~\ref{fig3}. From left to right: the relative phase between the first synchronous cluster and the 6th reference oscillator oscillates, whereas the phase of the second synchronous cluster passes zero and rotates until the clusters exchange their roles. Supplementary Movie 4 details this process. Parameters are $N=11$, $\mu=1.0$, $\varepsilon_1=1.0$, $\alpha_1=1.7$, $\varepsilon_2=0.08$, $\alpha_2=-2.0$.}
\label{fig6}
\end{figure}

\begin{figure}[h!]\center
\includegraphics[width=0.8\columnwidth]{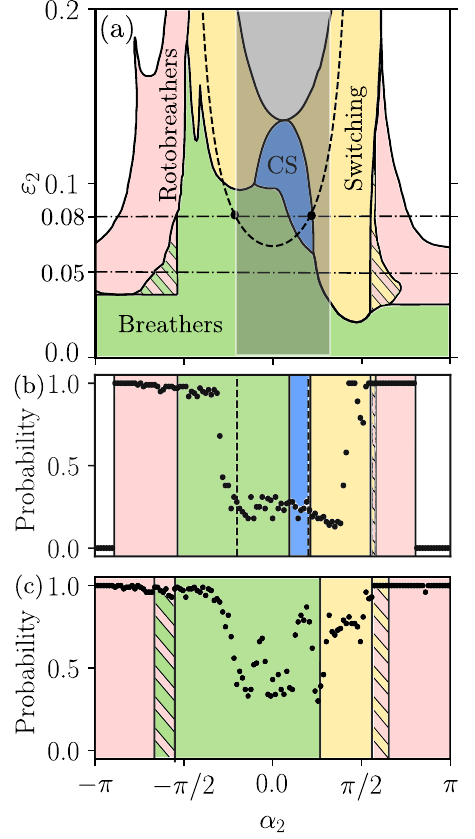}
	\caption{Stability and prevalence of cyclops states. (a). Stability diagram extending Fig.~\ref{fig1}a to the full range of the phase lag parameter $\alpha_2.$ The notations are similar to Fig.~\ref{fig1}a, with the addition of rotobreathers (pink). The shaded vertical strip corresponds to the parameter region of Fig.~\ref{fig1}a. Stationary cyclops states in the region $CS$ are chosen as initial conditions and further continued by changing the parameter $\alpha_2$ right and left from each point on the line $\alpha_2=0.0$ for each value of $\varepsilon_2.$ The initial conditions for the subsequent calculation are carried over from the final state of the preceding computation. The double-shaded areas (inclined stripes) indicate overlapping stability regions and correspond to the bistability of different cyclops state types. The two dash-dotted horizontal lines indicate the values of $\varepsilon_2$ used in panels b and c. (b, c). Probability of cyclops states' emergence (all types). The number of trials is $1,000.$  The initial phases are uniformly distributed in the segment $[-\pi, \pi]$, and the initial velocities are uniformly distributed in the segment $[-1.0, 1.0]$. The black dashed vertical lines in panel b indicate the stability boundary of full synchronization. In panel c, full synchronization is unstable.
    Parameters are $N=11$, $\mu=1.0$, $\varepsilon_1=1.0$, $\alpha_1=1.7$. (b): $\varepsilon_2=0.08$, (c): $\varepsilon_2=0.05$.}
\label{fig5}
\end{figure}

Breathing and switching cyclops states can also merge to form another hybrid cyclops state, termed {\it rotobreathing cyclops states}  (Fig.~\ref{fig6}) in the range of the second-harmonic phase shift with $|\alpha_2|>\pi/2$ (Fig.~\ref{fig5}, the pink regions). Rotobreathing cyclops states, or simply rotobreathers, are also characterized by a  two-stage repetitive process in which, during the first stage,
an inter-cluster phase difference between one cluster and the solitary oscillator oscillates while the relative phase difference of the other cluster rotates. The clusters exchange their oscillatory and rotatory phase roles during the second stage.
Figure~\ref{fig6} and Supplementary Movie 4 give the full details of this two-stage process. Accordingly, 
the amplitudes of the order parameters $r_1$ and $r_2$ exhibit large periodic oscillations  (Fig.~\ref{fig6}b). 

Figure~\ref{fig5} demonstrates the prevalence of 
cyclops states of various types. Remarkably, rotobreathers and breathing cyclops states, induced by non-zero second-harmonic phase lag $\alpha_2$ in the region where full synchronization is unstable, act as global attractors and emerge with a probability close to 1 (Fig.~\ref{fig5} b,c). Note that breathing and switching cyclops states can also emerge with a relatively high probability even when they co-exist with presumably dominant full synchronization when the overall coupling is attractive with $H'(0)>0$ (the region bounded by the black dashed vertical lines in Fig.~\ref{fig5}b; these lines correspond to the solid circles on the black dashed parabola in Fig.~\ref{fig5}a).

%\begin{figure}[h!]\center
%\includegraphics[width=1.0\columnwidth]{fig/Fig5.pdf}
%	\caption{Cyclops regimes diagram for system~\eqref{eq:system_origin}. Stationary cyclops states (phase differences $x$ and $y$ are constant) are shown in blue color, breather cyclops states (phase differences $x$ and $y$ are periodic, but without phase slips on $2\pi$) -- green color, rotobreather cyclops states (phase differences $x$ and $y$ are rotate and experience oscillaions relative to the rotational component of the movement) -- red color, switching cyclops states (part of the time there is a cyclops regime, where phase differences $x$ and $y$ are constant or periodic without phase slips on $2\pi$; other part of the time the cyclops clusters are rearranged) -- yellow color, other regimes -- white color. The dotted line indicates the boundaries of the area depicted in Fig.~\ref{fig1}a. The initial conditions are chosen by continuously changing the parameters $\varepsilon_2$ along the line $\alpha_2=0.0$ (the initial value of the next calculation is the final state of the system~\eqref{eq:system_origin} in the previous calculation) from the area $CS$, and after independent changing the parameters $\alpha_2$ right (a) and left (b) from each point of the line $\alpha_2=0.0$. \textcolor{red}{The black dotted line indicates the stability boundary of the fully synchronous mode ($H'(0)=0$). Fully synchronous mode is stable above the curve.}Parameters: $N=11$, $m=1.0$, $\varepsilon_1=1.0$, $\alpha_1=1.7$.}
%\label{fig5}
%\end{figure}
\begin{figure}[b!]\center
\includegraphics[width=0.90\columnwidth]{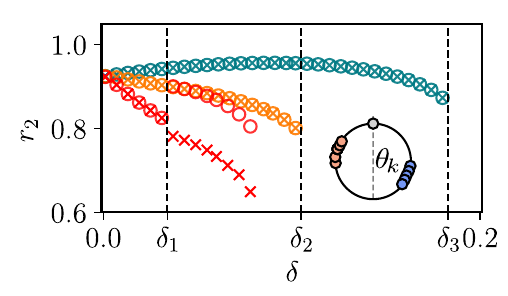}
	\caption{Persistence of cyclops states in system~\eqref{eq:system_origin} with mismatched frequencies $\omega_k$ distributed evenly over the interval  $[\omega - \delta,\; \omega + \delta],$ where $\omega=1.7$ and $\delta$ is a  frequency detuning. Global maxima (circles) and minima (crosses) of order parameter $r_2$ for three cyclops states (red, orange, and cyan). The values of $\delta < \delta_1$ preserve all three stable stationary cyclops states. Increasing $\delta > \delta_1$ destabilizes the first stationary cyclops state (red) and turns it into a switching cyclops state.  Further increasing $\delta > \delta_2$ leads to disintegrating the second cyclops state (orange) at $\delta=\delta_2$. The third stationary cyclops state (cyan) persists to $\delta=\delta_3$. The cyclops states are found from direct numerical simulations of system \eqref{eq:system_origin} for three sets of natural frequency distributions $\omega_k$ with a continuous increase in $\delta$ from zero. The inset shows instantaneous phase distributions $\theta_k$ for the third cyclops state with nonidentical frequencies. Parameters are $N=11$, $\mu=1.0$, $\varepsilon_1=1.0$, $\alpha_1=1.8$, $\varepsilon_2=0.12$, $\alpha_2=0.2$, $\delta_1=0.034$, $\delta_2=0.105$, $\delta_3=0.183$.}
\label{fig7}
\end{figure}

\section{Persistence of cyclops states}
In this section, we demonstrate that cyclops states resist intrinsic frequency detuning. We mismatch the  intrinsic frequency $\omega$ by choosing  
the $k$-th oscillator's frequency $\omega_k,$ $k=1,...N$ from a uniform random distribution in the interval $[\omega - \delta, \omega + \delta],$ where $\delta$ is a frequency detuning.
We consider the parameter region where stationary cyclops states are stable (region $CS$ in Fig.~\ref{fig1}a). Figure~\ref{fig7} demonstrates the persistence of three stationary cyclops states, each 
induced by a particular intrinsic frequency distribution.  Note that although the oscillators' phases within each synchronous cluster may not perfectly align due to the frequency detuning, they remain relatively close to each other (see Fig.~\ref{fig7}). Additionally, the established frequencies of all oscillators are the same.  The stationary cyclops state can lose the transversal stability similarly to their counterparts from the identical oscillator case (note the stationary cyclops state marked by the red labels in Fig.~\ref{fig7} that turns into a switching cyclops state at $\delta=\delta_1$). Remarkably, the frequency detuning can also induce a bifurcation scenario for disintegrating stationary cyclops states via a saddle-node bifurcation at $\delta=\delta_2$ and  $\delta=\delta_3.$ We did not observe such a  bifurcation route in our extensive simulations of system \eqref{eq:system_origin} with identical frequencies reported in Figs.~\ref{fig1}-\ref{fig5}.

%\begin{figure}[b!]\center
%\includegraphics[width=1.0\columnwidth]{fig/fig9.pdf}
%	\caption{\textcolor{red}{Probability (proportion of observations) of cyclops states (all types) realization calculated on the basis of $100$ experiments for each point. Initial conditions: the phases are uniformly distributed on the segment $[-\pi, \pi]$, the velocities are uniformly distributed on the segment $[-1.0, 1.0]$. The black dotted lines on fragment (a) indicate the stability boundary of the fully synchronous mode. On fragment (b) the fully synchronous regime is unstable. Parameters: $N=11$, $m=1.0$, $\varepsilon_1=1.0$, $\alpha_1=1.7$ and $\varepsilon_2=0.08$ (a), $\varepsilon_2=0.05$ (b).}}
%\label{fig8}
%\end{figure}

\section{Conclusions} 
Building upon our recent study \cite{munyayev2023cyclops}, 
this work has significantly advanced an understanding of rhythmogenesis in Kuramoto networks of 2D phase oscillators with first-mode and higher-mode coupling. A key focus of our work has been on the constructive role of higher coupling modes in inducing and stabilizing a unique class of dynamical states known as cyclops states. These states, characterized by two coherent clusters and a solitary oscillator resembling the Cyclops's eye, represent a particular form of three-cluster generalized splay states \cite{berner2021generalized}.

Our initial findings in \cite{munyayev2023cyclops} revealed the unexpected result that adding the second or third-harmonic to the Kuramoto coupling makes cyclops states global attractors, exhibiting remarkable stability over a substantial range of coupling's repulsion. This paper delved deeper into the dynamic repertoire of cyclops states, introducing and systematically analyzing breathing and switching cyclops states and their hybrids, including switching-breathing cyclops states and rotobreathers. Through rigorous analytical derivations and numerics, we have identified conditions for the existence and stability of stationary cyclops states, elucidating two distinct bifurcation scenarios. In both scenarios, the second coupling harmonic acts as a constructive agent, either inducing periodic oscillations in inter-cluster relative phase differences (breathing cyclops states) or facilitating swift reconfigurations and transitions (switching cyclops states).
These novel dynamical patterns can be viewed as nontrivial hybrids of solitary states \cite{maistrenko2017smallest,jaros2018solitary,teichmann2019solitary,munyayev2022stability}, generalized splay \cite{berner2021generalized}, clusters with breathing and rotatory inter-cluster phase shifts \cite{belykh2016bistability,brister2020three},
and intermittent \cite{olmi2015intermittent} and switching chimeras \cite{goldschmidt2019blinking}. In particular, switching cyclops states unite the properties of 
blinking chimeras \cite{goldschmidt2019blinking} and three-cluster states \cite{brister2020three}. 

Our extensive stability analysis has underscored the resilience and dominance of breathing, roto-breathing, and switching cyclops states across wide parameter ranges, including the case of the overall attractive, two-harmonic coupling.  Importantly, we have showcased that the constructive influence of higher coupling harmonics is not limited to networks of identical oscillators, as cyclops states persist robustly in Kuramoto networks of non-identical oscillators. 

 Importantly, our prior work \cite{munyayev2023cyclops} demonstrated the dynamic equivalence of the 2D Kuramoto model with first and second-harmonic coupling to a network of canonical theta-neurons with adaptive coupling. This equivalence also suggests the widespread manifestation of breathing and switching cyclops states in theta-neuron networks, underscoring our results' broad applicability and significance in diverse physical and biological networks.

While it is crucial to differentiate between the higher-order harmonic coupling studied in this paper and the higher-order non-pairwise coupling \cite{gambuzza2021stability,xu2021spectrum,millan2020explosive,boccaletti2023structure}, it is equally important to recognize their possible interplay and the richness they bring to the dynamics of networked systems. These concepts are not mutually exclusive; they can coexist, adding layers of complexity and fostering a diverse range of emergent behaviors. Recent research \cite{carballosa2023cluster,jaros2023higher} analyzed the intricate interplay between pairwise first-order harmonic and non-pairwise higher-order coupling in shaping collective dynamics in Kuramoto networks. The incorporation of both higher-order harmonics and non-pairwise interactions promises to induce even richer 
emerging dynamics, including various forms of cyclops states, and may pave the way for a more holistic comprehension of complex networked systems.

\begin{acknowledgments}
This work was supported by the Ministry of Science
and Higher Education of the Russian Federation under
project No.~0729-2020-0036 (to G.\,V.\,O.), the Russian
Science Foundation under project No.~22-12-00348 (to
M.\,I.\,B., V.\,O.\,M. and L.\,A.\,S.), the National Science
Foundation (USA) under Grant No. CMMI-2009329, and
the Office of Naval Research under Grant No. N00014-
22-1-2200 (to I.\,B.)
\end{acknowledgments}

\appendix
\section{Maximum number of stationary cyclops states}
Here, we provide the details for deriving an upper bound for the maximum number of stationary cyclops states with distinct $x$ and $y,$ given in Sec.~\ref{possible}.

Finding all possible solutions of system \eqref{eq:cyclope_find_nonlin} that determine the existence of stationary cyclops states is elusive due to its complexity, and the number of solutions can vary depending on the parameters. In particular, it prevents locating all solutions of the system \eqref{eq:cyclope_find_nonlin} by their continuation with respect to the parameters. However, this computational problem can be simplified by the change of variables $u=e^{ix}$, $v=e^{iy}$ ($|u|=1$, $|v|=1$) that transforms the real-valued  system \eqref{eq:cyclope_find_nonlin} into the system of complex 
 polynomial equations:
\begin{equation}
    \begin{aligned}
        (1-u)\Big(uv[&e^{i\alpha_1} (u+v+\frac{2uv}{N-1})\\
        &+e^{-i\alpha_1} v(u+v+\frac{2}{N-1})]\\
        &+\varepsilon_2 (u+1)[e^{i\alpha_2} (u^2+v^2+\frac{2u^2v^2}{N-1})\\
        &+e^{-i\alpha_2} v^2 (u^2+v^2+\frac{2}{N-1})]\Big)=0,\\
        (1-v)\Big(uv[&e^{i\alpha_1} (u+v+\frac{2uv}{N-1})\\
        &+e^{-i\alpha_1} u(u+v+\frac{2}{N-1})]\\
        &+\varepsilon_2 (v+1)[e^{i\alpha_2} (u^2+v^2+\frac{2u^2v^2}{N-1})\\
        &+e^{-i\alpha_2} u^2 (u^2+v^2+\frac{2}{N-1})]\Big)=0.
    \end{aligned}
    \label{eq:cyclope_find_lin}
\end{equation}
The analysis of system~\eqref{eq:cyclope_find_lin} is more manageable, and the maximum number of its solutions (the roots of the complex polynomials) can be estimated by applying the classical Bernshtein theorem from algebra. To facilitate the reading, we list this theorem below.\\
{\bf Theorem} [Bernshtein, 1975] \cite{bernshtein1975number}.
{\it Let a system of $n$ polynomials have a finite number of roots in $\bigl(C^* \bigr)^n$, where $C^*=C \setminus 0$. Then, the number of roots is bounded from above by the mixed volume $P_k$ of their Newton polytopes (the convex hull of polynomial supports $S_k$).}

Before applying the theorem to   \eqref{eq:cyclope_find_lin}, we get rid of the factors $(1-u)$ and $(1-v)$ on the right-hand side of \eqref{eq:cyclope_find_lin} since we are only interested in solutions $u, v \ne 1.$ By doing so, we have excluded 
the solutions that correspond to a one-cluster solution and two-cluster solutions of the form $(N-1)/2 : (N+1)/2.$ It is worth mentioning that, in contrast to its real-valued counterpart \eqref{eq:cyclope_find_nonlin}, the complex polynomials may have either non-physical solutions with $|u| \ne 1$ or $|v| \ne 1$, or solutions that do not correspond to stationary cyclops states. The latter solutions with $|u|=1$ and $|v|=1,$ include a two-cluster $N-1:1$ solitary state, corresponding to $u=v,$ 
i.e., $x=y$. 

\begin{figure}[b!]\center
\includegraphics[width=0.95\columnwidth]{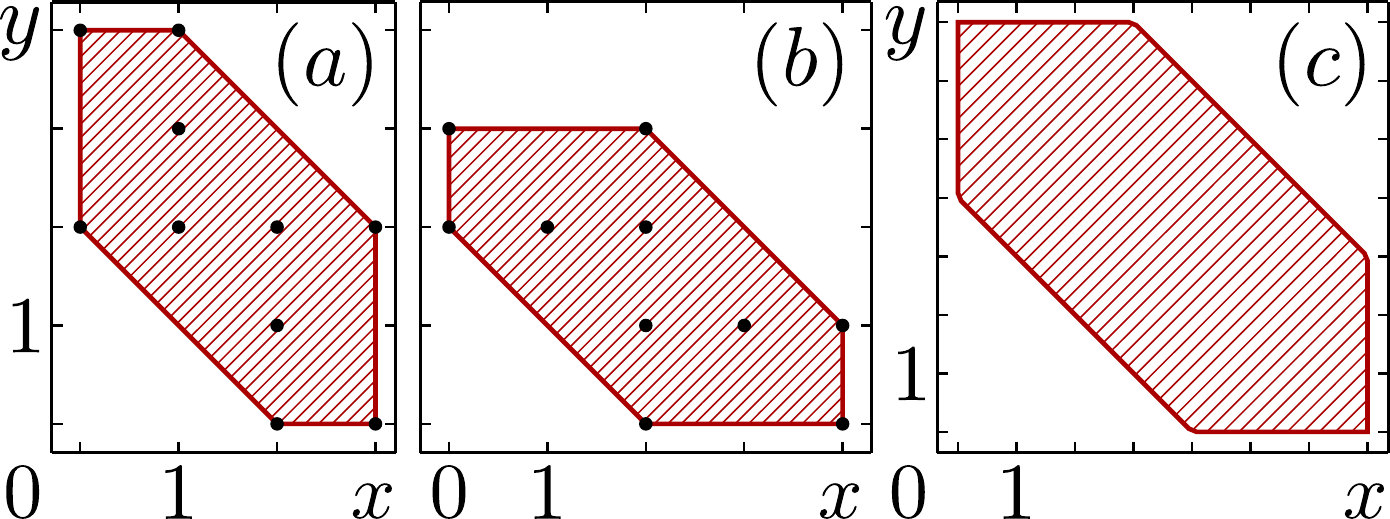}
	\caption{The supports $S_1$, $S_2$ (black dots) and the corresponding Newton polytopes $P_1$, $P_2$ (shaded regions) of (a) the first and (b) second polynomials of system \eqref{eq:cyclope_find_lin}. (c) The Minkowski sum $P_1\oplus P_2$.}
\label{fig:polytopes}
\end{figure}

The supports of the resulting polynomials (degrees $u$ and $v$ in each of the first and second equations terms) have the form: $S_1=\{(0, 2); (0, 4); (1, 2); (1, 3); (1, 4); (2, 0); (2, 1); (2, 2); (3, 0); \\(3, 2) \}$, $S_2=\{(0, 2); (0, 3); (1, 2); (2, 0); (2, 1); (2, 2); (2, 3);\\ (3, 1); (4, 0); (4, 1)\}$. Consider the mixed volume of Newtonian polytopes $P_1$ and $P_2$: $M(P_1, P_2) = \text{vol}_2 (P_1 \oplus P_2) - \text{vol}_2(P_1) - \text{vol}_2(P_2)$. The mixed volumes have the following values: $\text{vol}_2(P_1)=8$, $\text{vol}_2(P_2) = 8$, $\text{vol}_2 (P_1 \oplus P_2) = 33$ (see Fig.~\ref{fig:polytopes}). Therefore, $M(P_1, P_2) = 17$. Excluding the non-physical solutions and solutions corresponding to non-cyclops regimes from the sets of roots in \eqref{eq:cyclope_find_lin}, we can always find the number of cyclops modes in the system~\eqref{eq:system_origin} which is limited to $16$ cyclops states.

Our numerical search for the roots of polynomials \eqref{eq:cyclope_find_lin} was performed using the NSolve function of Wolfram Mathematica. This search found $17$ roots almost everywhere in the considered broad parameter regions, suggesting that our analysis effectively identified all possible solutions of\eqref{eq:cyclope_find_lin} and, therefore, all possible stationary states cyclops, identified from the 17 solutions by excluding the non-physical solutions ($|u| \ne 1$ or $|v| \ne 1$) and non-cyclops states ({$u=1$ or $v=1$ or $u=v$).

%\bibliography{references_breathing_cyclopes}
%

%\bibliography{references_cyclope_states}

\begin{thebibliography}{0}%
\makeatletter
\providecommand \@ifxundefined [1]{%
 \@ifx{#1\undefined}
}%
\providecommand \@ifnum [1]{%
 \ifnum #1\expandafter \@firstoftwo
 \else \expandafter \@secondoftwo
 \fi
}%
\providecommand \@ifx [1]{%
 \ifx #1\expandafter \@firstoftwo
 \else \expandafter \@secondoftwo
 \fi
}%
\providecommand \natexlab [1]{#1}%
\providecommand \enquote  [1]{``#1''}%
\providecommand \bibnamefont  [1]{#1}%
\providecommand \bibfnamefont [1]{#1}%
\providecommand \citenamefont [1]{#1}%
\providecommand \href@noop [0]{\@secondoftwo}%
\providecommand \href [0]{\begingroup \@sanitize@url \@href}%
\providecommand \@href[1]{\@@startlink{#1}\@@href}%
\providecommand \@@href[1]{\endgroup#1\@@endlink}%
\providecommand \@sanitize@url [0]{\catcode `\\12\catcode `\$12\catcode
  `\&12\catcode `\#12\catcode `\^12\catcode `\_12\catcode `\%12\relax}%
\providecommand \@@startlink[1]{}%
\providecommand \@@endlink[0]{}%
\providecommand \url  [0]{\begingroup\@sanitize@url \@url }%
\providecommand \@url [1]{\endgroup\@href {#1}{\urlprefix }}%
\providecommand \urlprefix  [0]{URL }%
\providecommand \Eprint [0]{\href }%
\providecommand \doibase [0]{https://doi.org/}%
\providecommand \selectlanguage [0]{\@gobble}%
\providecommand \bibinfo  [0]{\@secondoftwo}%
\providecommand \bibfield  [0]{\@secondoftwo}%
\providecommand \translation [1]{[#1]}%
\providecommand \BibitemOpen [0]{}%
\providecommand \bibitemStop [0]{}%
\providecommand \bibitemNoStop [0]{.\EOS\space}%
\providecommand \EOS [0]{\spacefactor3000\relax}%
\providecommand \BibitemShut  [1]{\csname bibitem#1\endcsname}%
\let\auto@bib@innerbib\@empty
%</preamble>
\end{thebibliography}%


\begin{thebibliography}{73}%
\makeatletter
\providecommand \@ifxundefined [1]{%
 \@ifx{#1\undefined}
}%
\providecommand \@ifnum [1]{%
 \ifnum #1\expandafter \@firstoftwo
 \else \expandafter \@secondoftwo
 \fi
}%
\providecommand \@ifx [1]{%
 \ifx #1\expandafter \@firstoftwo
 \else \expandafter \@secondoftwo
 \fi
}%
\providecommand \natexlab [1]{#1}%
\providecommand \enquote  [1]{``#1''}%
\providecommand \bibnamefont  [1]{#1}%
\providecommand \bibfnamefont [1]{#1}%
\providecommand \citenamefont [1]{#1}%
\providecommand \href@noop [0]{\@secondoftwo}%
\providecommand \href [0]{\begingroup \@sanitize@url \@href}%
\providecommand \@href[1]{\@@startlink{#1}\@@href}%
\providecommand \@@href[1]{\endgroup#1\@@endlink}%
\providecommand \@sanitize@url [0]{\catcode `\\12\catcode `\$12\catcode
  `\&12\catcode `\#12\catcode `\^12\catcode `\_12\catcode `\%12\relax}%
\providecommand \@@startlink[1]{}%
\providecommand \@@endlink[0]{}%
\providecommand \url  [0]{\begingroup\@sanitize@url \@url }%
\providecommand \@url [1]{\endgroup\@href {#1}{\urlprefix }}%
\providecommand \urlprefix  [0]{URL }%
\providecommand \Eprint [0]{\href }%
\providecommand \doibase [0]{http://dx.doi.org/}%
\providecommand \selectlanguage [0]{\@gobble}%
\providecommand \bibinfo  [0]{\@secondoftwo}%
\providecommand \bibfield  [0]{\@secondoftwo}%
\providecommand \translation [1]{[#1]}%
\providecommand \BibitemOpen [0]{}%
\providecommand \bibitemStop [0]{}%
\providecommand \bibitemNoStop [0]{.\EOS\space}%
\providecommand \EOS [0]{\spacefactor3000\relax}%
\providecommand \BibitemShut  [1]{\csname bibitem#1\endcsname}%
\let\auto@bib@innerbib\@empty
%</preamble>
\bibitem [{\citenamefont {Rinzel}\ and\ \citenamefont
  {Ermentrout}(1998)}]{rinzel1998analysis}%
  \BibitemOpen
  \bibfield  {author} {\bibinfo {author} {\bibfnamefont {J.}~\bibnamefont
  {Rinzel}}\ and\ \bibinfo {author} {\bibfnamefont {G.~B.}\ \bibnamefont
  {Ermentrout}},\ }\href@noop {} {\bibfield  {journal} {\bibinfo  {journal}
  {Methods in neuronal modeling}\ }\textbf {\bibinfo {volume} {2}},\ \bibinfo
  {pages} {251} (\bibinfo {year} {1998})}\BibitemShut {NoStop}%
\bibitem [{\citenamefont {Ermentrout}\ and\ \citenamefont
  {Kleinfeld}(2001)}]{ermentrout2001traveling}%
  \BibitemOpen
  \bibfield  {author} {\bibinfo {author} {\bibfnamefont {G.~B.}\ \bibnamefont
  {Ermentrout}}\ and\ \bibinfo {author} {\bibfnamefont {D.}~\bibnamefont
  {Kleinfeld}},\ }\href@noop {} {\bibfield  {journal} {\bibinfo  {journal}
  {Neuron}\ }\textbf {\bibinfo {volume} {29}},\ \bibinfo {pages} {33} (\bibinfo
  {year} {2001})}\BibitemShut {NoStop}%
\bibitem [{\citenamefont {Hoppensteadt}\ and\ \citenamefont
  {Izhikevich}(2012)}]{hoppensteadt2012weakly}%
  \BibitemOpen
  \bibfield  {author} {\bibinfo {author} {\bibfnamefont {F.~C.}\ \bibnamefont
  {Hoppensteadt}}\ and\ \bibinfo {author} {\bibfnamefont {E.~M.}\ \bibnamefont
  {Izhikevich}},\ }\href@noop {} {\emph {\bibinfo {title} {Weakly connected
  neural networks}}},\ Vol.\ \bibinfo {volume} {126}\ (\bibinfo  {publisher}
  {Springer Science \& Business Media},\ \bibinfo {year} {2012})\BibitemShut
  {NoStop}%
\bibitem [{\citenamefont {Kozyreff}\ \emph {et~al.}(2000)\citenamefont
  {Kozyreff}, \citenamefont {Vladimirov},\ and\ \citenamefont
  {Mandel}}]{kozyreff2000global}%
  \BibitemOpen
  \bibfield  {author} {\bibinfo {author} {\bibfnamefont {G.}~\bibnamefont
  {Kozyreff}}, \bibinfo {author} {\bibfnamefont {A.}~\bibnamefont
  {Vladimirov}}, \ and\ \bibinfo {author} {\bibfnamefont {P.}~\bibnamefont
  {Mandel}},\ }\href@noop {} {\bibfield  {journal} {\bibinfo  {journal}
  {Physical Review Letters}\ }\textbf {\bibinfo {volume} {85}},\ \bibinfo
  {pages} {3809} (\bibinfo {year} {2000})}\BibitemShut {NoStop}%
\bibitem [{\citenamefont {Ding}\ \emph {et~al.}(2019)\citenamefont {Ding},
  \citenamefont {Belykh}, \citenamefont {Marandi},\ and\ \citenamefont
  {Miri}}]{ding2019dispersive}%
  \BibitemOpen
  \bibfield  {author} {\bibinfo {author} {\bibfnamefont {J.}~\bibnamefont
  {Ding}}, \bibinfo {author} {\bibfnamefont {I.}~\bibnamefont {Belykh}},
  \bibinfo {author} {\bibfnamefont {A.}~\bibnamefont {Marandi}}, \ and\
  \bibinfo {author} {\bibfnamefont {M.-A.}\ \bibnamefont {Miri}},\ }\href@noop
  {} {\bibfield  {journal} {\bibinfo  {journal} {Physical Review Applied}\
  }\textbf {\bibinfo {volume} {12}},\ \bibinfo {pages} {054039} (\bibinfo
  {year} {2019})}\BibitemShut {NoStop}%
\bibitem [{\citenamefont {Nair}\ \emph {et~al.}(2021)\citenamefont {Nair},
  \citenamefont {Hu}, \citenamefont {Berrill}, \citenamefont {Wiesenfeld},\
  and\ \citenamefont {Braiman}}]{nair2021using}%
  \BibitemOpen
  \bibfield  {author} {\bibinfo {author} {\bibfnamefont {N.}~\bibnamefont
  {Nair}}, \bibinfo {author} {\bibfnamefont {K.}~\bibnamefont {Hu}}, \bibinfo
  {author} {\bibfnamefont {M.}~\bibnamefont {Berrill}}, \bibinfo {author}
  {\bibfnamefont {K.}~\bibnamefont {Wiesenfeld}}, \ and\ \bibinfo {author}
  {\bibfnamefont {Y.}~\bibnamefont {Braiman}},\ }\href@noop {} {\bibfield
  {journal} {\bibinfo  {journal} {Physical Review Letters}\ }\textbf {\bibinfo
  {volume} {127}},\ \bibinfo {pages} {173901} (\bibinfo {year}
  {2021})}\BibitemShut {NoStop}%
\bibitem [{\citenamefont {Motter}\ \emph {et~al.}(2013)\citenamefont {Motter},
  \citenamefont {Myers}, \citenamefont {Anghel},\ and\ \citenamefont
  {Nishikawa}}]{motter2013spontaneous}%
  \BibitemOpen
  \bibfield  {author} {\bibinfo {author} {\bibfnamefont {A.~E.}\ \bibnamefont
  {Motter}}, \bibinfo {author} {\bibfnamefont {S.~A.}\ \bibnamefont {Myers}},
  \bibinfo {author} {\bibfnamefont {M.}~\bibnamefont {Anghel}}, \ and\ \bibinfo
  {author} {\bibfnamefont {T.}~\bibnamefont {Nishikawa}},\ }\href@noop {}
  {\bibfield  {journal} {\bibinfo  {journal} {Nature Physics}\ }\textbf
  {\bibinfo {volume} {9}},\ \bibinfo {pages} {191} (\bibinfo {year}
  {2013})}\BibitemShut {NoStop}%
\bibitem [{\citenamefont {D{\"o}rfler}\ \emph {et~al.}(2013)\citenamefont
  {D{\"o}rfler}, \citenamefont {Chertkov},\ and\ \citenamefont
  {Bullo}}]{dorfler2013synchronization}%
  \BibitemOpen
  \bibfield  {author} {\bibinfo {author} {\bibfnamefont {F.}~\bibnamefont
  {D{\"o}rfler}}, \bibinfo {author} {\bibfnamefont {M.}~\bibnamefont
  {Chertkov}}, \ and\ \bibinfo {author} {\bibfnamefont {F.}~\bibnamefont
  {Bullo}},\ }\href@noop {} {\bibfield  {journal} {\bibinfo  {journal}
  {Proceedings of the National Academy of Sciences}\ }\textbf {\bibinfo
  {volume} {110}},\ \bibinfo {pages} {2005} (\bibinfo {year}
  {2013})}\BibitemShut {NoStop}%
\bibitem [{\citenamefont {Berner}\ \emph
  {et~al.}(2021{\natexlab{a}})\citenamefont {Berner}, \citenamefont {Yanchuk},\
  and\ \citenamefont {Sch{\"o}ll}}]{berner2021adaptive}%
  \BibitemOpen
  \bibfield  {author} {\bibinfo {author} {\bibfnamefont {R.}~\bibnamefont
  {Berner}}, \bibinfo {author} {\bibfnamefont {S.}~\bibnamefont {Yanchuk}}, \
  and\ \bibinfo {author} {\bibfnamefont {E.}~\bibnamefont {Sch{\"o}ll}},\
  }\href@noop {} {\bibfield  {journal} {\bibinfo  {journal} {Physical Review
  E}\ }\textbf {\bibinfo {volume} {103}},\ \bibinfo {pages} {042315} (\bibinfo
  {year} {2021}{\natexlab{a}})}\BibitemShut {NoStop}%
\bibitem [{\citenamefont {Kuramoto}(1975)}]{kuramoto1975self}%
  \BibitemOpen
  \bibfield  {author} {\bibinfo {author} {\bibfnamefont {Y.}~\bibnamefont
  {Kuramoto}},\ }in\ \href@noop {} {\emph {\bibinfo {booktitle} {International
  Symposium on Mathematical Problems in Theoretical Physics}}}\ (\bibinfo
  {organization} {Springer},\ \bibinfo {year} {1975})\ pp.\ \bibinfo {pages}
  {420--422}\BibitemShut {NoStop}%
\bibitem [{\citenamefont {Strogatz}(2000)}]{strogatz2000kuramoto}%
  \BibitemOpen
  \bibfield  {author} {\bibinfo {author} {\bibfnamefont {S.~H.}\ \bibnamefont
  {Strogatz}},\ }\href@noop {} {\bibfield  {journal} {\bibinfo  {journal}
  {Physica D: Nonlinear Phenomena}\ }\textbf {\bibinfo {volume} {143}},\
  \bibinfo {pages} {1} (\bibinfo {year} {2000})}\BibitemShut {NoStop}%
\bibitem [{\citenamefont {Ermentrout}(1997)}]{ermentrout}%
  \BibitemOpen
  \bibfield  {author} {\bibinfo {author} {\bibfnamefont {B.}~\bibnamefont
  {Ermentrout}},\ }\href@noop {} {\bibfield  {journal} {\bibinfo  {journal}
  {Journal of Mathematical Biology}\ } (\bibinfo {year} {1997})}\BibitemShut
  {NoStop}%
\bibitem [{\citenamefont {Acebr{\'o}n}\ \emph {et~al.}(2005)\citenamefont
  {Acebr{\'o}n}, \citenamefont {Bonilla}, \citenamefont {Vicente},
  \citenamefont {Ritort},\ and\ \citenamefont {Spigler}}]{acebron}%
  \BibitemOpen
  \bibfield  {author} {\bibinfo {author} {\bibfnamefont {J.~A.}\ \bibnamefont
  {Acebr{\'o}n}}, \bibinfo {author} {\bibfnamefont {L.~L.}\ \bibnamefont
  {Bonilla}}, \bibinfo {author} {\bibfnamefont {C.~J.~P.}\ \bibnamefont
  {Vicente}}, \bibinfo {author} {\bibfnamefont {F.}~\bibnamefont {Ritort}}, \
  and\ \bibinfo {author} {\bibfnamefont {R.}~\bibnamefont {Spigler}},\
  }\href@noop {} {\bibfield  {journal} {\bibinfo  {journal} {Reviews of Modern
  Physics}\ }\textbf {\bibinfo {volume} {77}},\ \bibinfo {pages} {137}
  (\bibinfo {year} {2005})}\BibitemShut {NoStop}%
\bibitem [{\citenamefont {Barreto}\ \emph {et~al.}(2008)\citenamefont
  {Barreto}, \citenamefont {Hunt}, \citenamefont {Ott},\ and\ \citenamefont
  {So}}]{barreto2008synchronization}%
  \BibitemOpen
  \bibfield  {author} {\bibinfo {author} {\bibfnamefont {E.}~\bibnamefont
  {Barreto}}, \bibinfo {author} {\bibfnamefont {B.}~\bibnamefont {Hunt}},
  \bibinfo {author} {\bibfnamefont {E.}~\bibnamefont {Ott}}, \ and\ \bibinfo
  {author} {\bibfnamefont {P.}~\bibnamefont {So}},\ }\href@noop {} {\bibfield
  {journal} {\bibinfo  {journal} {Physical Review E}\ }\textbf {\bibinfo
  {volume} {77}},\ \bibinfo {pages} {036107} (\bibinfo {year}
  {2008})}\BibitemShut {NoStop}%
\bibitem [{\citenamefont {Ott}\ and\ \citenamefont
  {Antonsen}(2008)}]{ott2008low}%
  \BibitemOpen
  \bibfield  {author} {\bibinfo {author} {\bibfnamefont {E.}~\bibnamefont
  {Ott}}\ and\ \bibinfo {author} {\bibfnamefont {T.~M.}\ \bibnamefont
  {Antonsen}},\ }\href@noop {} {\bibfield  {journal} {\bibinfo  {journal}
  {Chaos: An Interdisciplinary Journal of Nonlinear Science}\ }\textbf
  {\bibinfo {volume} {18}},\ \bibinfo {pages} {037113} (\bibinfo {year}
  {2008})}\BibitemShut {NoStop}%
\bibitem [{\citenamefont {Hong}\ \emph {et~al.}(2007)\citenamefont {Hong},
  \citenamefont {Chat{\'e}}, \citenamefont {Park},\ and\ \citenamefont
  {Tang}}]{hong2007entrainment}%
  \BibitemOpen
  \bibfield  {author} {\bibinfo {author} {\bibfnamefont {H.}~\bibnamefont
  {Hong}}, \bibinfo {author} {\bibfnamefont {H.}~\bibnamefont {Chat{\'e}}},
  \bibinfo {author} {\bibfnamefont {H.}~\bibnamefont {Park}}, \ and\ \bibinfo
  {author} {\bibfnamefont {L.-H.}\ \bibnamefont {Tang}},\ }\href@noop {}
  {\bibfield  {journal} {\bibinfo  {journal} {Physical Review Letters}\
  }\textbf {\bibinfo {volume} {99}},\ \bibinfo {pages} {184101} (\bibinfo
  {year} {2007})}\BibitemShut {NoStop}%
\bibitem [{\citenamefont {Pikovsky}\ and\ \citenamefont
  {Rosenblum}(2008)}]{pikovsky2008partially}%
  \BibitemOpen
  \bibfield  {author} {\bibinfo {author} {\bibfnamefont {A.}~\bibnamefont
  {Pikovsky}}\ and\ \bibinfo {author} {\bibfnamefont {M.}~\bibnamefont
  {Rosenblum}},\ }\href@noop {} {\bibfield  {journal} {\bibinfo  {journal}
  {Physical Review Letters}\ }\textbf {\bibinfo {volume} {101}},\ \bibinfo
  {pages} {264103} (\bibinfo {year} {2008})}\BibitemShut {NoStop}%
\bibitem [{\citenamefont {Maistrenko}\ \emph {et~al.}(2004)\citenamefont
  {Maistrenko}, \citenamefont {Popovych}, \citenamefont {Burylko},\ and\
  \citenamefont {Tass}}]{maistrenko2004mechanism}%
  \BibitemOpen
  \bibfield  {author} {\bibinfo {author} {\bibfnamefont {Y.}~\bibnamefont
  {Maistrenko}}, \bibinfo {author} {\bibfnamefont {O.}~\bibnamefont
  {Popovych}}, \bibinfo {author} {\bibfnamefont {O.}~\bibnamefont {Burylko}}, \
  and\ \bibinfo {author} {\bibfnamefont {P.}~\bibnamefont {Tass}},\ }\href@noop
  {} {\bibfield  {journal} {\bibinfo  {journal} {Physical Review Letters}\
  }\textbf {\bibinfo {volume} {93}},\ \bibinfo {pages} {084102} (\bibinfo
  {year} {2004})}\BibitemShut {NoStop}%
\bibitem [{\citenamefont {D{\"o}rfler}\ and\ \citenamefont
  {Bullo}(2011)}]{dorfler2011critical}%
  \BibitemOpen
  \bibfield  {author} {\bibinfo {author} {\bibfnamefont {F.}~\bibnamefont
  {D{\"o}rfler}}\ and\ \bibinfo {author} {\bibfnamefont {F.}~\bibnamefont
  {Bullo}},\ }\href@noop {} {\bibfield  {journal} {\bibinfo  {journal} {SIAM
  Journal on Applied Dynamical Systems}\ }\textbf {\bibinfo {volume} {10}},\
  \bibinfo {pages} {1070} (\bibinfo {year} {2011})}\BibitemShut {NoStop}%
\bibitem [{\citenamefont {Tanaka}\ \emph
  {et~al.}(1997{\natexlab{a}})\citenamefont {Tanaka}, \citenamefont
  {Lichtenberg},\ and\ \citenamefont {Oishi}}]{tanaka1997first}%
  \BibitemOpen
  \bibfield  {author} {\bibinfo {author} {\bibfnamefont {H.-A.}\ \bibnamefont
  {Tanaka}}, \bibinfo {author} {\bibfnamefont {A.~J.}\ \bibnamefont
  {Lichtenberg}}, \ and\ \bibinfo {author} {\bibfnamefont {S.}~\bibnamefont
  {Oishi}},\ }\href@noop {} {\bibfield  {journal} {\bibinfo  {journal}
  {Physical Review Letters}\ }\textbf {\bibinfo {volume} {78}},\ \bibinfo
  {pages} {2104} (\bibinfo {year} {1997}{\natexlab{a}})}\BibitemShut {NoStop}%
\bibitem [{\citenamefont {Tanaka}\ \emph
  {et~al.}(1997{\natexlab{b}})\citenamefont {Tanaka}, \citenamefont
  {Lichtenberg},\ and\ \citenamefont {Oishi}}]{tanaka1997self}%
  \BibitemOpen
  \bibfield  {author} {\bibinfo {author} {\bibfnamefont {H.-A.}\ \bibnamefont
  {Tanaka}}, \bibinfo {author} {\bibfnamefont {A.~J.}\ \bibnamefont
  {Lichtenberg}}, \ and\ \bibinfo {author} {\bibfnamefont {S.}~\bibnamefont
  {Oishi}},\ }\href@noop {} {\bibfield  {journal} {\bibinfo  {journal} {Physica
  D: Nonlinear Phenomena}\ }\textbf {\bibinfo {volume} {100}},\ \bibinfo
  {pages} {279} (\bibinfo {year} {1997}{\natexlab{b}})}\BibitemShut {NoStop}%
\bibitem [{\citenamefont {Ji}\ \emph {et~al.}(2014)\citenamefont {Ji},
  \citenamefont {Peron}, \citenamefont {Rodrigues},\ and\ \citenamefont
  {Kurths}}]{ji2014low}%
  \BibitemOpen
  \bibfield  {author} {\bibinfo {author} {\bibfnamefont {P.}~\bibnamefont
  {Ji}}, \bibinfo {author} {\bibfnamefont {T.~K.}\ \bibnamefont {Peron}},
  \bibinfo {author} {\bibfnamefont {F.~A.}\ \bibnamefont {Rodrigues}}, \ and\
  \bibinfo {author} {\bibfnamefont {J.}~\bibnamefont {Kurths}},\ }\href@noop {}
  {\bibfield  {journal} {\bibinfo  {journal} {Scientific Reports}\ }\textbf
  {\bibinfo {volume} {4}} (\bibinfo {year} {2014})}\BibitemShut {NoStop}%
\bibitem [{\citenamefont {Munyaev}\ \emph {et~al.}(2020)\citenamefont
  {Munyaev}, \citenamefont {Smirnov}, \citenamefont {Kostin}, \citenamefont
  {Osipov},\ and\ \citenamefont {Pikovsky}}]{munyaev2020analytical}%
  \BibitemOpen
  \bibfield  {author} {\bibinfo {author} {\bibfnamefont {V.}~\bibnamefont
  {Munyaev}}, \bibinfo {author} {\bibfnamefont {L.}~\bibnamefont {Smirnov}},
  \bibinfo {author} {\bibfnamefont {V.}~\bibnamefont {Kostin}}, \bibinfo
  {author} {\bibfnamefont {G.}~\bibnamefont {Osipov}}, \ and\ \bibinfo {author}
  {\bibfnamefont {A.}~\bibnamefont {Pikovsky}},\ }\href@noop {} {\bibfield
  {journal} {\bibinfo  {journal} {New Journal of Physics}\ }\textbf {\bibinfo
  {volume} {22}},\ \bibinfo {pages} {023036} (\bibinfo {year}
  {2020})}\BibitemShut {NoStop}%
\bibitem [{\citenamefont {Komarov}\ \emph {et~al.}(2014)\citenamefont
  {Komarov}, \citenamefont {Gupta},\ and\ \citenamefont
  {Pikovsky}}]{komarov2014synchronization}%
  \BibitemOpen
  \bibfield  {author} {\bibinfo {author} {\bibfnamefont {M.}~\bibnamefont
  {Komarov}}, \bibinfo {author} {\bibfnamefont {S.}~\bibnamefont {Gupta}}, \
  and\ \bibinfo {author} {\bibfnamefont {A.}~\bibnamefont {Pikovsky}},\
  }\href@noop {} {\bibfield  {journal} {\bibinfo  {journal} {EPL (Europhysics
  Letters)}\ }\textbf {\bibinfo {volume} {106}},\ \bibinfo {pages} {40003}
  (\bibinfo {year} {2014})}\BibitemShut {NoStop}%
\bibitem [{\citenamefont {Martens}\ \emph {et~al.}(2009)\citenamefont
  {Martens}, \citenamefont {Barreto}, \citenamefont {Strogatz}, \citenamefont
  {Ott}, \citenamefont {So},\ and\ \citenamefont
  {Antonsen}}]{martens2009exact}%
  \BibitemOpen
  \bibfield  {author} {\bibinfo {author} {\bibfnamefont {E.~A.}\ \bibnamefont
  {Martens}}, \bibinfo {author} {\bibfnamefont {E.}~\bibnamefont {Barreto}},
  \bibinfo {author} {\bibfnamefont {S.}~\bibnamefont {Strogatz}}, \bibinfo
  {author} {\bibfnamefont {E.}~\bibnamefont {Ott}}, \bibinfo {author}
  {\bibfnamefont {P.}~\bibnamefont {So}}, \ and\ \bibinfo {author}
  {\bibfnamefont {T.}~\bibnamefont {Antonsen}},\ }\href@noop {} {\bibfield
  {journal} {\bibinfo  {journal} {Physical Review E}\ }\textbf {\bibinfo
  {volume} {79}},\ \bibinfo {pages} {026204} (\bibinfo {year}
  {2009})}\BibitemShut {NoStop}%
\bibitem [{\citenamefont {Barabash}\ \emph {et~al.}(2021)\citenamefont
  {Barabash}, \citenamefont {Belykh}, \citenamefont {Osipov},\ and\
  \citenamefont {Belykh}}]{barabash2021partial}%
  \BibitemOpen
  \bibfield  {author} {\bibinfo {author} {\bibfnamefont {N.~V.}\ \bibnamefont
  {Barabash}}, \bibinfo {author} {\bibfnamefont {V.~N.}\ \bibnamefont
  {Belykh}}, \bibinfo {author} {\bibfnamefont {G.~V.}\ \bibnamefont {Osipov}},
  \ and\ \bibinfo {author} {\bibfnamefont {I.~V.}\ \bibnamefont {Belykh}},\
  }\href@noop {} {\bibfield  {journal} {\bibinfo  {journal} {Chaos}\ }\textbf
  {\bibinfo {volume} {31}} (\bibinfo {year} {2021})}\BibitemShut {NoStop}%
\bibitem [{\citenamefont {G{\'o}mez-Gardenes}\ \emph
  {et~al.}(2011)\citenamefont {G{\'o}mez-Gardenes}, \citenamefont {G{\'o}mez},
  \citenamefont {Arenas},\ and\ \citenamefont {Moreno}}]{gomez2011explosive}%
  \BibitemOpen
  \bibfield  {author} {\bibinfo {author} {\bibfnamefont {J.}~\bibnamefont
  {G{\'o}mez-Gardenes}}, \bibinfo {author} {\bibfnamefont {S.}~\bibnamefont
  {G{\'o}mez}}, \bibinfo {author} {\bibfnamefont {A.}~\bibnamefont {Arenas}}, \
  and\ \bibinfo {author} {\bibfnamefont {Y.}~\bibnamefont {Moreno}},\
  }\href@noop {} {\bibfield  {journal} {\bibinfo  {journal} {Physical Review
  Letters}\ }\textbf {\bibinfo {volume} {106}},\ \bibinfo {pages} {128701}
  (\bibinfo {year} {2011})}\BibitemShut {NoStop}%
\bibitem [{\citenamefont {Ji}\ \emph {et~al.}(2013)\citenamefont {Ji},
  \citenamefont {Peron}, \citenamefont {Menck}, \citenamefont {Rodrigues},\
  and\ \citenamefont {Kurths}}]{ji2013cluster}%
  \BibitemOpen
  \bibfield  {author} {\bibinfo {author} {\bibfnamefont {P.}~\bibnamefont
  {Ji}}, \bibinfo {author} {\bibfnamefont {T.~K.~D.}\ \bibnamefont {Peron}},
  \bibinfo {author} {\bibfnamefont {P.~J.}\ \bibnamefont {Menck}}, \bibinfo
  {author} {\bibfnamefont {F.~A.}\ \bibnamefont {Rodrigues}}, \ and\ \bibinfo
  {author} {\bibfnamefont {J.}~\bibnamefont {Kurths}},\ }\href@noop {}
  {\bibfield  {journal} {\bibinfo  {journal} {Physical Review Letters}\
  }\textbf {\bibinfo {volume} {110}},\ \bibinfo {pages} {218701} (\bibinfo
  {year} {2013})}\BibitemShut {NoStop}%
\bibitem [{\citenamefont {Skardal}\ and\ \citenamefont
  {Arenas}(2014)}]{skardal2014disorder}%
  \BibitemOpen
  \bibfield  {author} {\bibinfo {author} {\bibfnamefont {P.~S.}\ \bibnamefont
  {Skardal}}\ and\ \bibinfo {author} {\bibfnamefont {A.}~\bibnamefont
  {Arenas}},\ }\href@noop {} {\bibfield  {journal} {\bibinfo  {journal}
  {Physical Review E}\ }\textbf {\bibinfo {volume} {89}},\ \bibinfo {pages}
  {062811} (\bibinfo {year} {2014})}\BibitemShut {NoStop}%
\bibitem [{\citenamefont {Nishikawa}\ and\ \citenamefont
  {Motter}(2016)}]{nishikawa2016symmetric}%
  \BibitemOpen
  \bibfield  {author} {\bibinfo {author} {\bibfnamefont {T.}~\bibnamefont
  {Nishikawa}}\ and\ \bibinfo {author} {\bibfnamefont {A.~E.}\ \bibnamefont
  {Motter}},\ }\href@noop {} {\bibfield  {journal} {\bibinfo  {journal}
  {Physical Review Letters}\ }\textbf {\bibinfo {volume} {117}},\ \bibinfo
  {pages} {114101} (\bibinfo {year} {2016})}\BibitemShut {NoStop}%
\bibitem [{\citenamefont {Nicolaou}\ \emph {et~al.}(2019)\citenamefont
  {Nicolaou}, \citenamefont {Eroglu},\ and\ \citenamefont
  {Motter}}]{nicolaou2019multifaceted}%
  \BibitemOpen
  \bibfield  {author} {\bibinfo {author} {\bibfnamefont {Z.~G.}\ \bibnamefont
  {Nicolaou}}, \bibinfo {author} {\bibfnamefont {D.}~\bibnamefont {Eroglu}}, \
  and\ \bibinfo {author} {\bibfnamefont {A.~E.}\ \bibnamefont {Motter}},\
  }\href@noop {} {\bibfield  {journal} {\bibinfo  {journal} {Physical Review
  X}\ }\textbf {\bibinfo {volume} {9}},\ \bibinfo {pages} {011017} (\bibinfo
  {year} {2019})}\BibitemShut {NoStop}%
\bibitem [{\citenamefont {Kuramoto}\ and\ \citenamefont
  {Battogtokh}(2002)}]{kuramoto2002coexistence}%
  \BibitemOpen
  \bibfield  {author} {\bibinfo {author} {\bibfnamefont {Y.}~\bibnamefont
  {Kuramoto}}\ and\ \bibinfo {author} {\bibfnamefont {D.}~\bibnamefont
  {Battogtokh}},\ }\href@noop {} {\bibfield  {journal} {\bibinfo  {journal}
  {Nonlinear Phenomena in Complex Systems}\ }\textbf {\bibinfo {volume} {5}},\
  \bibinfo {pages} {380} (\bibinfo {year} {2002})}\BibitemShut {NoStop}%
\bibitem [{\citenamefont {Abrams}\ and\ \citenamefont
  {Strogatz}(2004)}]{abrams2004chimera}%
  \BibitemOpen
  \bibfield  {author} {\bibinfo {author} {\bibfnamefont {D.~M.}\ \bibnamefont
  {Abrams}}\ and\ \bibinfo {author} {\bibfnamefont {S.~H.}\ \bibnamefont
  {Strogatz}},\ }\href@noop {} {\bibfield  {journal} {\bibinfo  {journal}
  {Physical Review Letters}\ }\textbf {\bibinfo {volume} {93}},\ \bibinfo
  {pages} {174102} (\bibinfo {year} {2004})}\BibitemShut {NoStop}%
\bibitem [{\citenamefont {Abrams}\ \emph {et~al.}(2008)\citenamefont {Abrams},
  \citenamefont {Mirollo}, \citenamefont {Strogatz},\ and\ \citenamefont
  {Wiley}}]{abrams2008solvable}%
  \BibitemOpen
  \bibfield  {author} {\bibinfo {author} {\bibfnamefont {D.~M.}\ \bibnamefont
  {Abrams}}, \bibinfo {author} {\bibfnamefont {R.}~\bibnamefont {Mirollo}},
  \bibinfo {author} {\bibfnamefont {S.~H.}\ \bibnamefont {Strogatz}}, \ and\
  \bibinfo {author} {\bibfnamefont {D.~A.}\ \bibnamefont {Wiley}},\ }\href@noop
  {} {\bibfield  {journal} {\bibinfo  {journal} {Physical Review Letters}\
  }\textbf {\bibinfo {volume} {101}},\ \bibinfo {pages} {084103} (\bibinfo
  {year} {2008})}\BibitemShut {NoStop}%
\bibitem [{\citenamefont {Panaggio}\ and\ \citenamefont
  {Abrams}(2015)}]{panaggio2015chimera}%
  \BibitemOpen
  \bibfield  {author} {\bibinfo {author} {\bibfnamefont {M.~J.}\ \bibnamefont
  {Panaggio}}\ and\ \bibinfo {author} {\bibfnamefont {D.~M.}\ \bibnamefont
  {Abrams}},\ }\href@noop {} {\bibfield  {journal} {\bibinfo  {journal}
  {Nonlinearity}\ }\textbf {\bibinfo {volume} {28}},\ \bibinfo {pages} {R67}
  (\bibinfo {year} {2015})}\BibitemShut {NoStop}%
\bibitem [{\citenamefont {Zakharova}\ \emph {et~al.}(2014)\citenamefont
  {Zakharova}, \citenamefont {Kapeller},\ and\ \citenamefont
  {Sch{\"o}ll}}]{zakharova2014chimera}%
  \BibitemOpen
  \bibfield  {author} {\bibinfo {author} {\bibfnamefont {A.}~\bibnamefont
  {Zakharova}}, \bibinfo {author} {\bibfnamefont {M.}~\bibnamefont {Kapeller}},
  \ and\ \bibinfo {author} {\bibfnamefont {E.}~\bibnamefont {Sch{\"o}ll}},\
  }\href@noop {} {\bibfield  {journal} {\bibinfo  {journal} {Physical Review
  Letters}\ }\textbf {\bibinfo {volume} {112}},\ \bibinfo {pages} {154101}
  (\bibinfo {year} {2014})}\BibitemShut {NoStop}%
\bibitem [{\citenamefont {Olmi}\ \emph {et~al.}(2015)\citenamefont {Olmi},
  \citenamefont {Martens}, \citenamefont {Thutupalli},\ and\ \citenamefont
  {Torcini}}]{olmi2015intermittent}%
  \BibitemOpen
  \bibfield  {author} {\bibinfo {author} {\bibfnamefont {S.}~\bibnamefont
  {Olmi}}, \bibinfo {author} {\bibfnamefont {E.~A.}\ \bibnamefont {Martens}},
  \bibinfo {author} {\bibfnamefont {S.}~\bibnamefont {Thutupalli}}, \ and\
  \bibinfo {author} {\bibfnamefont {A.}~\bibnamefont {Torcini}},\ }\href@noop
  {} {\bibfield  {journal} {\bibinfo  {journal} {Physical Review E}\ }\textbf
  {\bibinfo {volume} {92}},\ \bibinfo {pages} {030901} (\bibinfo {year}
  {2015})}\BibitemShut {NoStop}%
\bibitem [{\citenamefont {Bolotov}\ \emph {et~al.}(2016)\citenamefont
  {Bolotov}, \citenamefont {Osipov},\ and\ \citenamefont
  {Pikovsky}}]{bolotov2016marginal}%
  \BibitemOpen
  \bibfield  {author} {\bibinfo {author} {\bibfnamefont {M.}~\bibnamefont
  {Bolotov}}, \bibinfo {author} {\bibfnamefont {G.}~\bibnamefont {Osipov}}, \
  and\ \bibinfo {author} {\bibfnamefont {A.}~\bibnamefont {Pikovsky}},\
  }\href@noop {} {\bibfield  {journal} {\bibinfo  {journal} {Physical Review
  E}\ }\textbf {\bibinfo {volume} {93}},\ \bibinfo {pages} {032202} (\bibinfo
  {year} {2016})}\BibitemShut {NoStop}%
\bibitem [{\citenamefont {Bolotov}\ \emph {et~al.}(2018)\citenamefont
  {Bolotov}, \citenamefont {Smirnov}, \citenamefont {Osipov},\ and\
  \citenamefont {Pikovsky}}]{bolotov2018simple}%
  \BibitemOpen
  \bibfield  {author} {\bibinfo {author} {\bibfnamefont {M.}~\bibnamefont
  {Bolotov}}, \bibinfo {author} {\bibfnamefont {L.}~\bibnamefont {Smirnov}},
  \bibinfo {author} {\bibfnamefont {G.}~\bibnamefont {Osipov}}, \ and\ \bibinfo
  {author} {\bibfnamefont {A.}~\bibnamefont {Pikovsky}},\ }\href@noop {}
  {\bibfield  {journal} {\bibinfo  {journal} {Chaos: An Interdisciplinary
  Journal of Nonlinear Science}\ }\textbf {\bibinfo {volume} {28}},\ \bibinfo
  {pages} {045101} (\bibinfo {year} {2018})}\BibitemShut {NoStop}%
\bibitem [{\citenamefont {Jaros}\ \emph {et~al.}(2015)\citenamefont {Jaros},
  \citenamefont {Maistrenko},\ and\ \citenamefont
  {Kapitaniak}}]{jaros2015chimera}%
  \BibitemOpen
  \bibfield  {author} {\bibinfo {author} {\bibfnamefont {P.}~\bibnamefont
  {Jaros}}, \bibinfo {author} {\bibfnamefont {Y.}~\bibnamefont {Maistrenko}}, \
  and\ \bibinfo {author} {\bibfnamefont {T.}~\bibnamefont {Kapitaniak}},\
  }\href@noop {} {\bibfield  {journal} {\bibinfo  {journal} {Physical Review
  E}\ }\textbf {\bibinfo {volume} {91}},\ \bibinfo {pages} {022907} (\bibinfo
  {year} {2015})}\BibitemShut {NoStop}%
\bibitem [{\citenamefont {Maistrenko}\ \emph {et~al.}(2017)\citenamefont
  {Maistrenko}, \citenamefont {Brezetsky}, \citenamefont {Jaros}, \citenamefont
  {Levchenko},\ and\ \citenamefont {Kapitaniak}}]{maistrenko2017smallest}%
  \BibitemOpen
  \bibfield  {author} {\bibinfo {author} {\bibfnamefont {Y.}~\bibnamefont
  {Maistrenko}}, \bibinfo {author} {\bibfnamefont {S.}~\bibnamefont
  {Brezetsky}}, \bibinfo {author} {\bibfnamefont {P.}~\bibnamefont {Jaros}},
  \bibinfo {author} {\bibfnamefont {R.}~\bibnamefont {Levchenko}}, \ and\
  \bibinfo {author} {\bibfnamefont {T.}~\bibnamefont {Kapitaniak}},\
  }\href@noop {} {\bibfield  {journal} {\bibinfo  {journal} {Physical Review
  E}\ }\textbf {\bibinfo {volume} {95}},\ \bibinfo {pages} {010203} (\bibinfo
  {year} {2017})}\BibitemShut {NoStop}%
\bibitem [{\citenamefont {Jaros}\ \emph {et~al.}(2018)\citenamefont {Jaros},
  \citenamefont {Brezetsky}, \citenamefont {Levchenko}, \citenamefont
  {Dudkowski}, \citenamefont {Kapitaniak},\ and\ \citenamefont
  {Maistrenko}}]{jaros2018solitary}%
  \BibitemOpen
  \bibfield  {author} {\bibinfo {author} {\bibfnamefont {P.}~\bibnamefont
  {Jaros}}, \bibinfo {author} {\bibfnamefont {S.}~\bibnamefont {Brezetsky}},
  \bibinfo {author} {\bibfnamefont {R.}~\bibnamefont {Levchenko}}, \bibinfo
  {author} {\bibfnamefont {D.}~\bibnamefont {Dudkowski}}, \bibinfo {author}
  {\bibfnamefont {T.}~\bibnamefont {Kapitaniak}}, \ and\ \bibinfo {author}
  {\bibfnamefont {Y.}~\bibnamefont {Maistrenko}},\ }\href@noop {} {\bibfield
  {journal} {\bibinfo  {journal} {Chaos: An Interdisciplinary Journal of
  Nonlinear Science}\ }\textbf {\bibinfo {volume} {28}},\ \bibinfo {pages}
  {011103} (\bibinfo {year} {2018})}\BibitemShut {NoStop}%
\bibitem [{\citenamefont {Teichmann}\ and\ \citenamefont
  {Rosenblum}(2019)}]{teichmann2019solitary}%
  \BibitemOpen
  \bibfield  {author} {\bibinfo {author} {\bibfnamefont {E.}~\bibnamefont
  {Teichmann}}\ and\ \bibinfo {author} {\bibfnamefont {M.}~\bibnamefont
  {Rosenblum}},\ }\href@noop {} {\bibfield  {journal} {\bibinfo  {journal}
  {Chaos: An Interdisciplinary Journal of Nonlinear Science}\ }\textbf
  {\bibinfo {volume} {29}},\ \bibinfo {pages} {093124} (\bibinfo {year}
  {2019})}\BibitemShut {NoStop}%
\bibitem [{\citenamefont {Munyayev}\ \emph {et~al.}(2022)\citenamefont
  {Munyayev}, \citenamefont {Bolotov}, \citenamefont {Smirnov}, \citenamefont
  {Osipov},\ and\ \citenamefont {Belykh}}]{munyayev2022stability}%
  \BibitemOpen
  \bibfield  {author} {\bibinfo {author} {\bibfnamefont {V.~O.}\ \bibnamefont
  {Munyayev}}, \bibinfo {author} {\bibfnamefont {M.~I.}\ \bibnamefont
  {Bolotov}}, \bibinfo {author} {\bibfnamefont {L.~A.}\ \bibnamefont
  {Smirnov}}, \bibinfo {author} {\bibfnamefont {G.~V.}\ \bibnamefont {Osipov}},
  \ and\ \bibinfo {author} {\bibfnamefont {I.~V.}\ \bibnamefont {Belykh}},\
  }\href@noop {} {\bibfield  {journal} {\bibinfo  {journal} {Physical Review
  E}\ }\textbf {\bibinfo {volume} {105}},\ \bibinfo {pages} {024203} (\bibinfo
  {year} {2022})}\BibitemShut {NoStop}%
\bibitem [{\citenamefont {Olmi}\ \emph {et~al.}(2014)\citenamefont {Olmi},
  \citenamefont {Navas}, \citenamefont {Boccaletti},\ and\ \citenamefont
  {Torcini}}]{olmi2014hysteretic}%
  \BibitemOpen
  \bibfield  {author} {\bibinfo {author} {\bibfnamefont {S.}~\bibnamefont
  {Olmi}}, \bibinfo {author} {\bibfnamefont {A.}~\bibnamefont {Navas}},
  \bibinfo {author} {\bibfnamefont {S.}~\bibnamefont {Boccaletti}}, \ and\
  \bibinfo {author} {\bibfnamefont {A.}~\bibnamefont {Torcini}},\ }\href@noop
  {} {\bibfield  {journal} {\bibinfo  {journal} {Physical Review E}\ }\textbf
  {\bibinfo {volume} {90}},\ \bibinfo {pages} {042905} (\bibinfo {year}
  {2014})}\BibitemShut {NoStop}%
\bibitem [{\citenamefont {Belykh}\ \emph {et~al.}(2016)\citenamefont {Belykh},
  \citenamefont {Brister},\ and\ \citenamefont
  {Belykh}}]{belykh2016bistability}%
  \BibitemOpen
  \bibfield  {author} {\bibinfo {author} {\bibfnamefont {I.~V.}\ \bibnamefont
  {Belykh}}, \bibinfo {author} {\bibfnamefont {B.~N.}\ \bibnamefont {Brister}},
  \ and\ \bibinfo {author} {\bibfnamefont {V.~N.}\ \bibnamefont {Belykh}},\
  }\href@noop {} {\bibfield  {journal} {\bibinfo  {journal} {Chaos: An
  Interdisciplinary Journal of Nonlinear Science}\ }\textbf {\bibinfo {volume}
  {26}},\ \bibinfo {pages} {094822} (\bibinfo {year} {2016})}\BibitemShut
  {NoStop}%
\bibitem [{\citenamefont {Brister}\ \emph {et~al.}(2020)\citenamefont
  {Brister}, \citenamefont {Belykh},\ and\ \citenamefont
  {Belykh}}]{brister2020three}%
  \BibitemOpen
  \bibfield  {author} {\bibinfo {author} {\bibfnamefont {B.~N.}\ \bibnamefont
  {Brister}}, \bibinfo {author} {\bibfnamefont {V.~N.}\ \bibnamefont {Belykh}},
  \ and\ \bibinfo {author} {\bibfnamefont {I.~V.}\ \bibnamefont {Belykh}},\
  }\href@noop {} {\bibfield  {journal} {\bibinfo  {journal} {Physical Review
  E}\ }\textbf {\bibinfo {volume} {101}},\ \bibinfo {pages} {062206} (\bibinfo
  {year} {2020})}\BibitemShut {NoStop}%
\bibitem [{\citenamefont {Ronge}\ and\ \citenamefont
  {Zaks}(2021)}]{ronge2021splay}%
  \BibitemOpen
  \bibfield  {author} {\bibinfo {author} {\bibfnamefont {R.}~\bibnamefont
  {Ronge}}\ and\ \bibinfo {author} {\bibfnamefont {M.~A.}\ \bibnamefont
  {Zaks}},\ }\href@noop {} {\bibfield  {journal} {\bibinfo  {journal} {The
  European Physical Journal Special Topics}\ }\textbf {\bibinfo {volume}
  {230}},\ \bibinfo {pages} {2717} (\bibinfo {year} {2021})}\BibitemShut
  {NoStop}%
\bibitem [{\citenamefont {Berner}\ \emph
  {et~al.}(2021{\natexlab{b}})\citenamefont {Berner}, \citenamefont {Yanchuk},
  \citenamefont {Maistrenko},\ and\ \citenamefont
  {Scholl}}]{berner2021generalized}%
  \BibitemOpen
  \bibfield  {author} {\bibinfo {author} {\bibfnamefont {R.}~\bibnamefont
  {Berner}}, \bibinfo {author} {\bibfnamefont {S.}~\bibnamefont {Yanchuk}},
  \bibinfo {author} {\bibfnamefont {Y.}~\bibnamefont {Maistrenko}}, \ and\
  \bibinfo {author} {\bibfnamefont {E.}~\bibnamefont {Scholl}},\ }\href@noop {}
  {\bibfield  {journal} {\bibinfo  {journal} {Chaos: An Interdisciplinary
  Journal of Nonlinear Science}\ }\textbf {\bibinfo {volume} {31}},\ \bibinfo
  {pages} {073128} (\bibinfo {year} {2021}{\natexlab{b}})}\BibitemShut
  {NoStop}%
\bibitem [{\citenamefont {Munyayev}\ \emph {et~al.}(2023)\citenamefont
  {Munyayev}, \citenamefont {Bolotov}, \citenamefont {Smirnov}, \citenamefont
  {Osipov},\ and\ \citenamefont {Belykh}}]{munyayev2023cyclops}%
  \BibitemOpen
  \bibfield  {author} {\bibinfo {author} {\bibfnamefont {V.~O.}\ \bibnamefont
  {Munyayev}}, \bibinfo {author} {\bibfnamefont {M.~I.}\ \bibnamefont
  {Bolotov}}, \bibinfo {author} {\bibfnamefont {L.~A.}\ \bibnamefont
  {Smirnov}}, \bibinfo {author} {\bibfnamefont {G.~V.}\ \bibnamefont {Osipov}},
  \ and\ \bibinfo {author} {\bibfnamefont {I.}~\bibnamefont {Belykh}},\
  }\href@noop {} {\bibfield  {journal} {\bibinfo  {journal} {Physical Review
  Letters}\ }\textbf {\bibinfo {volume} {130}},\ \bibinfo {pages} {107201}
  (\bibinfo {year} {2023})}\BibitemShut {NoStop}%
\bibitem [{\citenamefont {Tsimring}\ \emph {et~al.}(2005)\citenamefont
  {Tsimring}, \citenamefont {Rulkov}, \citenamefont {Larsen},\ and\
  \citenamefont {Gabbay}}]{tsimring2005repulsive}%
  \BibitemOpen
  \bibfield  {author} {\bibinfo {author} {\bibfnamefont {L.}~\bibnamefont
  {Tsimring}}, \bibinfo {author} {\bibfnamefont {N.}~\bibnamefont {Rulkov}},
  \bibinfo {author} {\bibfnamefont {M.}~\bibnamefont {Larsen}}, \ and\ \bibinfo
  {author} {\bibfnamefont {M.}~\bibnamefont {Gabbay}},\ }\href@noop {}
  {\bibfield  {journal} {\bibinfo  {journal} {Physical Review Letters}\
  }\textbf {\bibinfo {volume} {95}},\ \bibinfo {pages} {014101} (\bibinfo
  {year} {2005})}\BibitemShut {NoStop}%
\bibitem [{\citenamefont {Gao}\ \emph {et~al.}(2019)\citenamefont {Gao},
  \citenamefont {Fu}, \citenamefont {Cai}, \citenamefont {Yang},\ and\
  \citenamefont {Eugene~Stanley}}]{gao2019repulsive}%
  \BibitemOpen
  \bibfield  {author} {\bibinfo {author} {\bibfnamefont {Y.-C.}\ \bibnamefont
  {Gao}}, \bibinfo {author} {\bibfnamefont {C.-J.}\ \bibnamefont {Fu}},
  \bibinfo {author} {\bibfnamefont {S.-M.}\ \bibnamefont {Cai}}, \bibinfo
  {author} {\bibfnamefont {C.}~\bibnamefont {Yang}}, \ and\ \bibinfo {author}
  {\bibfnamefont {H.}~\bibnamefont {Eugene~Stanley}},\ }\href@noop {}
  {\bibfield  {journal} {\bibinfo  {journal} {Chaos: An Interdisciplinary
  Journal of Nonlinear Science}\ }\textbf {\bibinfo {volume} {29}},\ \bibinfo
  {pages} {053130} (\bibinfo {year} {2019})}\BibitemShut {NoStop}%
\bibitem [{\citenamefont {Belykh}\ and\ \citenamefont
  {Shilnikov}(2008)}]{belykh2008weak}%
  \BibitemOpen
  \bibfield  {author} {\bibinfo {author} {\bibfnamefont {I.}~\bibnamefont
  {Belykh}}\ and\ \bibinfo {author} {\bibfnamefont {A.}~\bibnamefont
  {Shilnikov}},\ }\href@noop {} {\bibfield  {journal} {\bibinfo  {journal}
  {Physical Review Letters}\ }\textbf {\bibinfo {volume} {101}},\ \bibinfo
  {pages} {078102} (\bibinfo {year} {2008})}\BibitemShut {NoStop}%
\bibitem [{\citenamefont {Nishikawa}\ and\ \citenamefont
  {Motter}(2010)}]{nishikawa2010network}%
  \BibitemOpen
  \bibfield  {author} {\bibinfo {author} {\bibfnamefont {T.}~\bibnamefont
  {Nishikawa}}\ and\ \bibinfo {author} {\bibfnamefont {A.~E.}\ \bibnamefont
  {Motter}},\ }\href@noop {} {\bibfield  {journal} {\bibinfo  {journal}
  {Proceedings of the National Academy of Sciences}\ }\textbf {\bibinfo
  {volume} {107}},\ \bibinfo {pages} {10342} (\bibinfo {year}
  {2010})}\BibitemShut {NoStop}%
\bibitem [{\citenamefont {Belykh}\ \emph {et~al.}(2015)\citenamefont {Belykh},
  \citenamefont {Reimbayev},\ and\ \citenamefont
  {Zhao}}]{belykh2015synergistic}%
  \BibitemOpen
  \bibfield  {author} {\bibinfo {author} {\bibfnamefont {I.}~\bibnamefont
  {Belykh}}, \bibinfo {author} {\bibfnamefont {R.}~\bibnamefont {Reimbayev}}, \
  and\ \bibinfo {author} {\bibfnamefont {K.}~\bibnamefont {Zhao}},\ }\href@noop
  {} {\bibfield  {journal} {\bibinfo  {journal} {Physical Review E}\ }\textbf
  {\bibinfo {volume} {91}},\ \bibinfo {pages} {062919} (\bibinfo {year}
  {2015})}\BibitemShut {NoStop}%
\bibitem [{\citenamefont {Reimbayev}\ \emph {et~al.}(2017)\citenamefont
  {Reimbayev}, \citenamefont {Daley},\ and\ \citenamefont
  {Belykh}}]{reimbayev2017two}%
  \BibitemOpen
  \bibfield  {author} {\bibinfo {author} {\bibfnamefont {R.}~\bibnamefont
  {Reimbayev}}, \bibinfo {author} {\bibfnamefont {K.}~\bibnamefont {Daley}}, \
  and\ \bibinfo {author} {\bibfnamefont {I.}~\bibnamefont {Belykh}},\
  }\href@noop {} {\bibfield  {journal} {\bibinfo  {journal} {Philosophical
  Transactions of the Royal Society A: Mathematical, Physical and Engineering
  Sciences}\ }\textbf {\bibinfo {volume} {375}},\ \bibinfo {pages} {20160282}
  (\bibinfo {year} {2017})}\BibitemShut {NoStop}%
\bibitem [{\citenamefont {Seliger}\ \emph {et~al.}(2002)\citenamefont
  {Seliger}, \citenamefont {Young},\ and\ \citenamefont
  {Tsimring}}]{seliger2002plasticity}%
  \BibitemOpen
  \bibfield  {author} {\bibinfo {author} {\bibfnamefont {P.}~\bibnamefont
  {Seliger}}, \bibinfo {author} {\bibfnamefont {S.~C.}\ \bibnamefont {Young}},
  \ and\ \bibinfo {author} {\bibfnamefont {L.~S.}\ \bibnamefont {Tsimring}},\
  }\href@noop {} {\bibfield  {journal} {\bibinfo  {journal} {Physical Review
  E}\ }\textbf {\bibinfo {volume} {65}},\ \bibinfo {pages} {041906} (\bibinfo
  {year} {2002})}\BibitemShut {NoStop}%
\bibitem [{\citenamefont {Niyogi}\ and\ \citenamefont
  {English}(2009)}]{niyogi2009learning}%
  \BibitemOpen
  \bibfield  {author} {\bibinfo {author} {\bibfnamefont {R.~K.}\ \bibnamefont
  {Niyogi}}\ and\ \bibinfo {author} {\bibfnamefont {L.~Q.}\ \bibnamefont
  {English}},\ }\href@noop {} {\bibfield  {journal} {\bibinfo  {journal}
  {Physical Review E}\ }\textbf {\bibinfo {volume} {80}},\ \bibinfo {pages}
  {066213} (\bibinfo {year} {2009})}\BibitemShut {NoStop}%
\bibitem [{\citenamefont {Kiss}\ \emph {et~al.}(2005)\citenamefont {Kiss},
  \citenamefont {Zhai},\ and\ \citenamefont {Hudson}}]{kiss2005predicting}%
  \BibitemOpen
  \bibfield  {author} {\bibinfo {author} {\bibfnamefont {I.~Z.}\ \bibnamefont
  {Kiss}}, \bibinfo {author} {\bibfnamefont {Y.}~\bibnamefont {Zhai}}, \ and\
  \bibinfo {author} {\bibfnamefont {J.~L.}\ \bibnamefont {Hudson}},\
  }\href@noop {} {\bibfield  {journal} {\bibinfo  {journal} {Physical Review
  Letters}\ }\textbf {\bibinfo {volume} {94}},\ \bibinfo {pages} {248301}
  (\bibinfo {year} {2005})}\BibitemShut {NoStop}%
\bibitem [{\citenamefont {Goldobin}\ \emph {et~al.}(2013)\citenamefont
  {Goldobin}, \citenamefont {Kleiner}, \citenamefont {Koelle},\ and\
  \citenamefont {Mints}}]{goldobin2013phase}%
  \BibitemOpen
  \bibfield  {author} {\bibinfo {author} {\bibfnamefont {E.}~\bibnamefont
  {Goldobin}}, \bibinfo {author} {\bibfnamefont {R.}~\bibnamefont {Kleiner}},
  \bibinfo {author} {\bibfnamefont {D.}~\bibnamefont {Koelle}}, \ and\ \bibinfo
  {author} {\bibfnamefont {R.}~\bibnamefont {Mints}},\ }\href@noop {}
  {\bibfield  {journal} {\bibinfo  {journal} {Physical Review Letters}\
  }\textbf {\bibinfo {volume} {111}},\ \bibinfo {pages} {057004} (\bibinfo
  {year} {2013})}\BibitemShut {NoStop}%
\bibitem [{\citenamefont {Komarov}\ and\ \citenamefont
  {Pikovsky}(2013)}]{komarov2013multiplicity}%
  \BibitemOpen
  \bibfield  {author} {\bibinfo {author} {\bibfnamefont {M.}~\bibnamefont
  {Komarov}}\ and\ \bibinfo {author} {\bibfnamefont {A.}~\bibnamefont
  {Pikovsky}},\ }\href@noop {} {\bibfield  {journal} {\bibinfo  {journal}
  {Physical Review Letters}\ }\textbf {\bibinfo {volume} {111}},\ \bibinfo
  {pages} {204101} (\bibinfo {year} {2013})}\BibitemShut {NoStop}%
\bibitem [{\citenamefont {Berner}\ \emph {et~al.}(2023)\citenamefont {Berner},
  \citenamefont {Lu},\ and\ \citenamefont
  {Sokolov}}]{berner2023synchronization}%
  \BibitemOpen
  \bibfield  {author} {\bibinfo {author} {\bibfnamefont {R.}~\bibnamefont
  {Berner}}, \bibinfo {author} {\bibfnamefont {A.}~\bibnamefont {Lu}}, \ and\
  \bibinfo {author} {\bibfnamefont {I.~M.}\ \bibnamefont {Sokolov}},\
  }\href@noop {} {\bibfield  {journal} {\bibinfo  {journal} {Chaos: An
  Interdisciplinary Journal of Nonlinear Science}\ }\textbf {\bibinfo {volume}
  {33}},\ \bibinfo {pages} {073138} (\bibinfo {year} {2023})}\BibitemShut
  {NoStop}%
\bibitem [{\citenamefont {Skardal}\ \emph {et~al.}(2011)\citenamefont
  {Skardal}, \citenamefont {Ott},\ and\ \citenamefont
  {Restrepo}}]{skardal2011cluster}%
  \BibitemOpen
  \bibfield  {author} {\bibinfo {author} {\bibfnamefont {P.~S.}\ \bibnamefont
  {Skardal}}, \bibinfo {author} {\bibfnamefont {E.}~\bibnamefont {Ott}}, \ and\
  \bibinfo {author} {\bibfnamefont {J.~G.}\ \bibnamefont {Restrepo}},\
  }\href@noop {} {\bibfield  {journal} {\bibinfo  {journal} {Physical Review
  E}\ }\textbf {\bibinfo {volume} {84}},\ \bibinfo {pages} {036208} (\bibinfo
  {year} {2011})}\BibitemShut {NoStop}%
\bibitem [{\citenamefont {Goldschmidt}\ \emph {et~al.}(2019)\citenamefont
  {Goldschmidt}, \citenamefont {Pikovsky},\ and\ \citenamefont
  {Politi}}]{goldschmidt2019blinking}%
  \BibitemOpen
  \bibfield  {author} {\bibinfo {author} {\bibfnamefont {R.~J.}\ \bibnamefont
  {Goldschmidt}}, \bibinfo {author} {\bibfnamefont {A.}~\bibnamefont
  {Pikovsky}}, \ and\ \bibinfo {author} {\bibfnamefont {A.}~\bibnamefont
  {Politi}},\ }\href@noop {} {\bibfield  {journal} {\bibinfo  {journal} {Chaos:
  An Interdisciplinary Journal of Nonlinear Science}\ }\textbf {\bibinfo
  {volume} {29}} (\bibinfo {year} {2019})}\BibitemShut {NoStop}%
\bibitem [{\citenamefont {Sakaguchi}(2006)}]{sakaguchi2006instability}%
  \BibitemOpen
  \bibfield  {author} {\bibinfo {author} {\bibfnamefont {H.}~\bibnamefont
  {Sakaguchi}},\ }\href@noop {} {\bibfield  {journal} {\bibinfo  {journal}
  {Physical Review E}\ }\textbf {\bibinfo {volume} {73}},\ \bibinfo {pages}
  {031907} (\bibinfo {year} {2006})}\BibitemShut {NoStop}%
\bibitem [{\citenamefont {Daido}(1992)}]{daido1992order}%
  \BibitemOpen
  \bibfield  {author} {\bibinfo {author} {\bibfnamefont {H.}~\bibnamefont
  {Daido}},\ }\href@noop {} {\bibfield  {journal} {\bibinfo  {journal}
  {Progress of Theoretical Physics}\ }\textbf {\bibinfo {volume} {88}},\
  \bibinfo {pages} {1213} (\bibinfo {year} {1992})}\BibitemShut {NoStop}%
\bibitem [{\citenamefont {Bernshtein}(1975)}]{bernshtein1975number}%
  \BibitemOpen
  \bibfield  {author} {\bibinfo {author} {\bibfnamefont {D.~N.}\ \bibnamefont
  {Bernshtein}},\ }\href@noop {} {\bibfield  {journal} {\bibinfo  {journal}
  {Funktsional'nyi Analiz i Ego Prilozheniya}\ }\textbf {\bibinfo {volume}
  {9}},\ \bibinfo {pages} {1} (\bibinfo {year} {1975})}\BibitemShut {NoStop}%
\bibitem [{\citenamefont {Gambuzza}\ \emph {et~al.}(2021)\citenamefont
  {Gambuzza}, \citenamefont {Di~Patti}, \citenamefont {Gallo}, \citenamefont
  {Lepri}, \citenamefont {Romance}, \citenamefont {Criado}, \citenamefont
  {Frasca}, \citenamefont {Latora},\ and\ \citenamefont
  {Boccaletti}}]{gambuzza2021stability}%
  \BibitemOpen
  \bibfield  {author} {\bibinfo {author} {\bibfnamefont {L.~V.}\ \bibnamefont
  {Gambuzza}}, \bibinfo {author} {\bibfnamefont {F.}~\bibnamefont {Di~Patti}},
  \bibinfo {author} {\bibfnamefont {L.}~\bibnamefont {Gallo}}, \bibinfo
  {author} {\bibfnamefont {S.}~\bibnamefont {Lepri}}, \bibinfo {author}
  {\bibfnamefont {M.}~\bibnamefont {Romance}}, \bibinfo {author} {\bibfnamefont
  {R.}~\bibnamefont {Criado}}, \bibinfo {author} {\bibfnamefont
  {M.}~\bibnamefont {Frasca}}, \bibinfo {author} {\bibfnamefont
  {V.}~\bibnamefont {Latora}}, \ and\ \bibinfo {author} {\bibfnamefont
  {S.}~\bibnamefont {Boccaletti}},\ }\href@noop {} {\bibfield  {journal}
  {\bibinfo  {journal} {Nature communications}\ }\textbf {\bibinfo {volume}
  {12}},\ \bibinfo {pages} {1255} (\bibinfo {year} {2021})}\BibitemShut
  {NoStop}%
\bibitem [{\citenamefont {Xu}\ and\ \citenamefont
  {Skardal}(2021)}]{xu2021spectrum}%
  \BibitemOpen
  \bibfield  {author} {\bibinfo {author} {\bibfnamefont {C.}~\bibnamefont
  {Xu}}\ and\ \bibinfo {author} {\bibfnamefont {P.~S.}\ \bibnamefont
  {Skardal}},\ }\href@noop {} {\bibfield  {journal} {\bibinfo  {journal}
  {Physical Review Research}\ }\textbf {\bibinfo {volume} {3}},\ \bibinfo
  {pages} {013013} (\bibinfo {year} {2021})}\BibitemShut {NoStop}%
\bibitem [{\citenamefont {Mill{\'a}n}\ \emph {et~al.}(2020)\citenamefont
  {Mill{\'a}n}, \citenamefont {Torres},\ and\ \citenamefont
  {Bianconi}}]{millan2020explosive}%
  \BibitemOpen
  \bibfield  {author} {\bibinfo {author} {\bibfnamefont {A.~P.}\ \bibnamefont
  {Mill{\'a}n}}, \bibinfo {author} {\bibfnamefont {J.~J.}\ \bibnamefont
  {Torres}}, \ and\ \bibinfo {author} {\bibfnamefont {G.}~\bibnamefont
  {Bianconi}},\ }\href@noop {} {\bibfield  {journal} {\bibinfo  {journal}
  {Physical Review Letters}\ }\textbf {\bibinfo {volume} {124}},\ \bibinfo
  {pages} {218301} (\bibinfo {year} {2020})}\BibitemShut {NoStop}%
\bibitem [{\citenamefont {Boccaletti}\ \emph {et~al.}(2023)\citenamefont
  {Boccaletti}, \citenamefont {De~Lellis}, \citenamefont {del Genio},
  \citenamefont {Alfaro-Bittner}, \citenamefont {Criado}, \citenamefont
  {Jalan},\ and\ \citenamefont {Romance}}]{boccaletti2023structure}%
  \BibitemOpen
  \bibfield  {author} {\bibinfo {author} {\bibfnamefont {S.}~\bibnamefont
  {Boccaletti}}, \bibinfo {author} {\bibfnamefont {P.}~\bibnamefont
  {De~Lellis}}, \bibinfo {author} {\bibfnamefont {C.}~\bibnamefont {del
  Genio}}, \bibinfo {author} {\bibfnamefont {K.}~\bibnamefont
  {Alfaro-Bittner}}, \bibinfo {author} {\bibfnamefont {R.}~\bibnamefont
  {Criado}}, \bibinfo {author} {\bibfnamefont {S.}~\bibnamefont {Jalan}}, \
  and\ \bibinfo {author} {\bibfnamefont {M.}~\bibnamefont {Romance}},\
  }\href@noop {} {\bibfield  {journal} {\bibinfo  {journal} {Physics Reports}\
  }\textbf {\bibinfo {volume} {1018}},\ \bibinfo {pages} {1} (\bibinfo {year}
  {2023})}\BibitemShut {NoStop}%
\bibitem [{\citenamefont {Carballosa}\ \emph {et~al.}(2023)\citenamefont
  {Carballosa}, \citenamefont {Mu{\~n}uzuri}, \citenamefont {Boccaletti},
  \citenamefont {Torcini},\ and\ \citenamefont {Olmi}}]{carballosa2023cluster}%
  \BibitemOpen
  \bibfield  {author} {\bibinfo {author} {\bibfnamefont {A.}~\bibnamefont
  {Carballosa}}, \bibinfo {author} {\bibfnamefont {A.~P.}\ \bibnamefont
  {Mu{\~n}uzuri}}, \bibinfo {author} {\bibfnamefont {S.}~\bibnamefont
  {Boccaletti}}, \bibinfo {author} {\bibfnamefont {A.}~\bibnamefont {Torcini}},
  \ and\ \bibinfo {author} {\bibfnamefont {S.}~\bibnamefont {Olmi}},\
  }\href@noop {} {\bibfield  {journal} {\bibinfo  {journal} {Chaos, Solitons \&
  Fractals}\ }\textbf {\bibinfo {volume} {177}},\ \bibinfo {pages} {114197}
  (\bibinfo {year} {2023})}\BibitemShut {NoStop}%
\bibitem [{\citenamefont {Jaros}\ \emph {et~al.}(2023)\citenamefont {Jaros},
  \citenamefont {Ghosh}, \citenamefont {Dudkowski}, \citenamefont {Dana},\ and\
  \citenamefont {Kapitaniak}}]{jaros2023higher}%
  \BibitemOpen
  \bibfield  {author} {\bibinfo {author} {\bibfnamefont {P.}~\bibnamefont
  {Jaros}}, \bibinfo {author} {\bibfnamefont {S.}~\bibnamefont {Ghosh}},
  \bibinfo {author} {\bibfnamefont {D.}~\bibnamefont {Dudkowski}}, \bibinfo
  {author} {\bibfnamefont {S.~K.}\ \bibnamefont {Dana}}, \ and\ \bibinfo
  {author} {\bibfnamefont {T.}~\bibnamefont {Kapitaniak}},\ }\href@noop {}
  {\bibfield  {journal} {\bibinfo  {journal} {Physical Review E}\ }\textbf
  {\bibinfo {volume} {108}},\ \bibinfo {pages} {024215} (\bibinfo {year}
  {2023})}\BibitemShut {NoStop}%
\end{thebibliography}

\end{document}